\documentclass[secnumarabic,amssymb, nobibnotes, aps, prx,reprint,preprintnumbers,superscriptaddress,amsmath,bibnotes,longbibliography]{revtex4-2}
\usepackage{graphicx}% Include figure files
\usepackage{dcolumn}% Align table columns on decimal point
\usepackage{bm}% bold math
\usepackage[colorlinks,linkcolor=blue,anchorcolor=blue,citecolor=blue,urlcolor=blue,filecolor=blue,menucolor=blue,runcolor=blue]{hyperref}% add hypertext capabilities

\begin{document}
\title{Dimensionality tuning of heavy-fermion states in ultrathin CeSi$_{2}$ films}

\author{Yi Wu}
\thanks{These authors contributed equally to this paper}
\affiliation{School of Physics, Zhejiang University, Hangzhou 310058, China}
\affiliation{New Cornerstone Science Laboratory and Center for Correlated Matter, Zhejiang University, Hangzhou 310058, China}
\author{Weifan Zhu}
\thanks{These authors contributed equally to this paper}
\affiliation{School of Physics, Zhejiang University, Hangzhou 310058, China}
\affiliation{New Cornerstone Science Laboratory and Center for Correlated Matter, Zhejiang University, Hangzhou 310058, China}
\author{Teng Hua}
\thanks{These authors contributed equally to this paper}
\affiliation{School of Physics, Zhejiang University, Hangzhou 310058, China}
\affiliation{New Cornerstone Science Laboratory and Center for Correlated Matter, Zhejiang University, Hangzhou 310058, China}
\author{Yuan Fang}
\thanks{These authors contributed equally to this paper}
\affiliation{Laboratory of Thin Film Optics, Shanghai Institute of Optics and Fine Mechanics, Chinese Academy of Sciences, Shanghai 201800, China}
\author{Yanan Zhang}
\affiliation{School of Physics, Zhejiang University, Hangzhou 310058, China}
\affiliation{New Cornerstone Science Laboratory and Center for Correlated Matter, Zhejiang University, Hangzhou 310058, China}
\author{Jiawen Zhang}
\affiliation{School of Physics, Zhejiang University, Hangzhou 310058, China}
\affiliation{New Cornerstone Science Laboratory and Center for Correlated Matter, Zhejiang University, Hangzhou 310058, China}
\author{Yanen Huang}
\affiliation{School of Physics, Zhejiang University, Hangzhou 310058, China}
\affiliation{New Cornerstone Science Laboratory and Center for Correlated Matter, Zhejiang University, Hangzhou 310058, China}
\author{Hao Zheng}
\affiliation{School of Physics, Zhejiang University, Hangzhou 310058, China}
\affiliation{New Cornerstone Science Laboratory and Center for Correlated Matter, Zhejiang University, Hangzhou 310058, China}
\author{{Shanyin Fu}}
\affiliation{School of Physics, Zhejiang University, Hangzhou 310058, China}
\affiliation{New Cornerstone Science Laboratory and Center for Correlated Matter, Zhejiang University, Hangzhou 310058, China}
\author{{Xinying Zheng}}
\affiliation{School of Physics, Zhejiang University, Hangzhou 310058, China}
\affiliation{New Cornerstone Science Laboratory and Center for Correlated Matter, Zhejiang University, Hangzhou 310058, China}
\author{{Zhengtai Liu}}
\affiliation{Shanghai Synchrotron Radiation Facility, Shanghai Advanced Research Institute, Chinese Academy of Sciences, Shanghai 201204, China}
\author{{Mao Ye}}
\affiliation{Shanghai Synchrotron Radiation Facility, Shanghai Advanced Research Institute, Chinese Academy of Sciences, Shanghai 201204, China}
\author{{Ye Chen}}
\affiliation{School of Physics, Zhejiang University, Hangzhou 310058, China}
\affiliation{New Cornerstone Science Laboratory and Center for Correlated Matter, Zhejiang University, Hangzhou 310058, China}
\author{Tulai Sun}
\thanks{Corresponding author: tlsun2020@zjut.edu.cn}
\affiliation{Center for Electron Microscopy, Zhejiang University of Technology, Hanghzhou 310014, China}
\author{Michael Smidman}
\affiliation{School of Physics, Zhejiang University, Hangzhou 310058, China}
\affiliation{New Cornerstone Science Laboratory and Center for Correlated Matter, Zhejiang University, Hangzhou 310058, China}
\author{Johann Kroha}
\affiliation{Physikalisches Institut and Bethe Center for Theoretical Physics, Universit\"at Bonn, Nussallee 12, 53115 Bonn, Germany}
\affiliation{School of Physics and Astronomy, University of St. Andrews, North Haugh, St. Andrews, KY16 9SS, United Kingdom}
\author{Chao Cao}
\affiliation{School of Physics, Zhejiang University, Hangzhou 310058, China}
\affiliation{New Cornerstone Science Laboratory and Center for Correlated Matter, Zhejiang University, Hangzhou 310058, China}
\affiliation{Institute for Advanced Study in Physics, Zhejiang University, Hangzhou 310058, China}
\author{Huiqiu Yuan}
\thanks{Corresponding author: hqyuan@zju.edu.cn}
\affiliation{School of Physics, Zhejiang University, Hangzhou 310058, China}
\affiliation{New Cornerstone Science Laboratory and Center for Correlated Matter, Zhejiang University, Hangzhou 310058, China}
\affiliation{Institute for Advanced Study in Physics, Zhejiang University, Hangzhou 310058, China}
\author{Frank Steglich}
\affiliation{School of Physics, Zhejiang University, Hangzhou 310058, China}
\affiliation{New Cornerstone Science Laboratory and Center for Correlated Matter, Zhejiang University, Hangzhou 310058, China}
\affiliation{Max Planck Institute for Chemical Physics of Solids, 01187 Dresden, Germany}
\author{Hai-Qing Lin}
\affiliation{School of Physics, Zhejiang University, Hangzhou 310058, China}
\affiliation{Institute for Advanced Study in Physics, Zhejiang University, Hangzhou 310058, China}
\author{Yang Liu}
\thanks{Corresponding author: yangliuphys@zju.edu.cn}
\affiliation{School of Physics, Zhejiang University, Hangzhou 310058, China}
\affiliation{New Cornerstone Science Laboratory and Center for Correlated Matter, Zhejiang University, Hangzhou 310058, China}
\affiliation{Institute for Advanced Study in Physics, Zhejiang University, Hangzhou 310058, China}
\date{\today}%
\addcontentsline{toc}{chapter}{Abstract}
%\author[1,5]{Yang Liu\thanks{Corresponding author: yangliuphys@zju.edu.cn}}

\begin{abstract}
Dimensionality tuning is an important method to modify the electronic states of quantum materials. However, the mechanism of such tuning in heavy fermion systems and its connection with transport properties remain largely unexplored. Here by combining molecular beam epitaxy (MBE), $in$-$situ$ angle-resolved photoemission spectroscopy (ARPES) and transport measurements, we study the electronic states of the heavy-fermion compound CeSi$_{2}$ as a function of film thickness. In three-dimensional thick films, our measurements reveal a dispersive Kondo peak at the Fermi level ($E_F$) and satellite peaks originating from crystal electric field (CEF) excitations, characteristic of heavy-fermion systems. For two-dimensional ultrathin films, the CEF satellites are largely suppressed while the ground-state Kondo peak at $E_F$ remains strong, although it develops at lower temperatures. Simultaneously, the maximum temperature $T_{max}$ of the magnetic resistivity, $\rho_m$($T$), changes from $\approx$100 K in thick films to $\approx$35 K in ultrathin films. This can be attributed to the dimensionality-driven reduction of CEF excitations during the Kondo process, in good agreement with spectroscopic results. Our work provides direct insight to understand the quantum confinement effects on strongly correlated 4$f$-electron systems and opens up new opportunities to explore emergent phenomena in two-dimensional heavy-fermion materials.
\end{abstract}

\maketitle

\section{INTRODUCTION}

\par Dimensionality is an important tuning parameter to modify the electronic states of quantum materials and plays an important role in exploring new quantum phenomena in condensed matter physics. Compared to their three-dimensional counterparts, the electronic wave functions in two-dimensional systems become confined within the $x-y$ plane and correspondingly stronger correlations among charge carriers and enhanced quantum fluctuations develop at low temperatures. This can lead to important phenomena, e.g., high-temperature superconductivity and fractional quantum Hall effect. Dimensionality tuning has been studied extensively by electron spectroscopy in itinerant $s$/$p$-electron systems \cite{ChiangPhotoemission}, as well as correlated $d$-electron systems \cite{Yoshimatsu2011Science,Monkman2012NM}, where quantum confinement effects and enhanced electron correlations in the two-dimensional limit have been well documented. In particular, in some strongly correlated $3d$-electron systems, largely reduced quasiparticle weight $Z$ has been observed in the two-dimensional limit \cite{King2014}. Nevertheless, dimensionality tuning in correlated $f$-electron systems has been much less explored by electron spectroscopy.

\begin{figure*}[ht]
    \includegraphics[width=1.6\columnwidth]{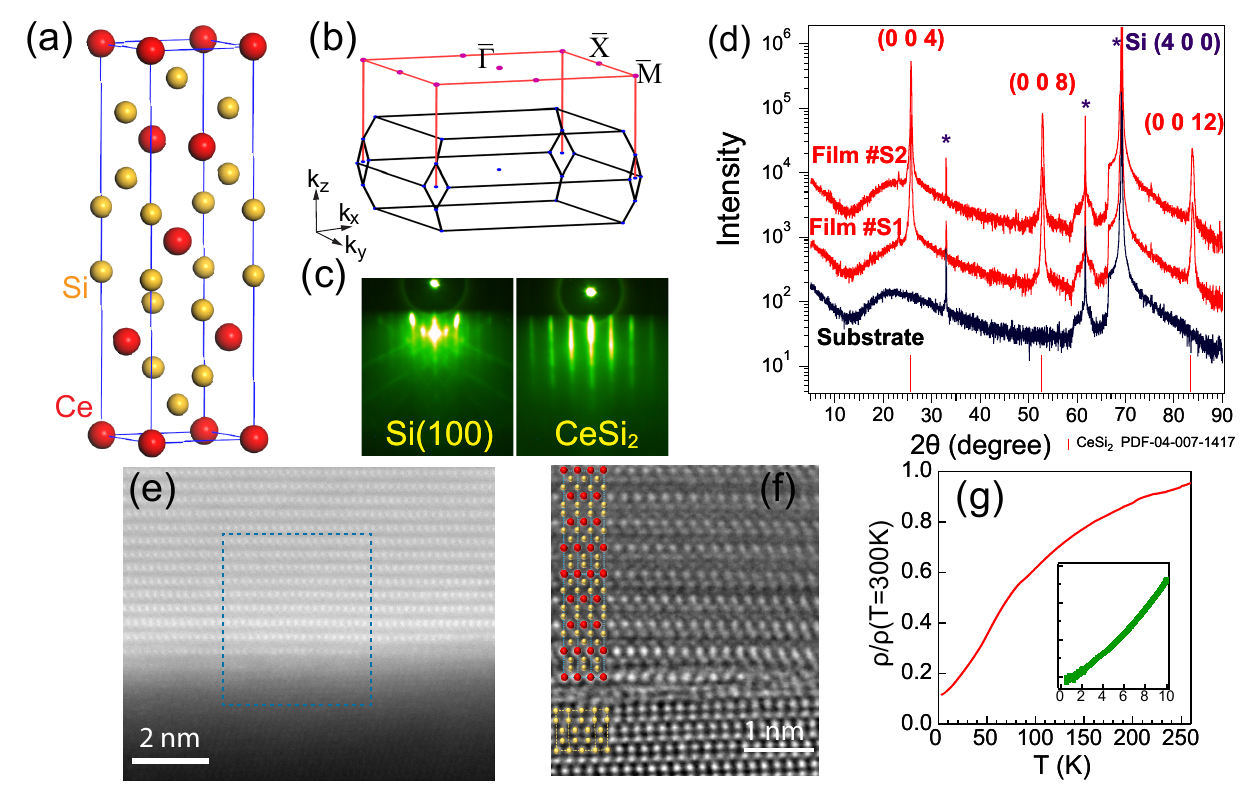}
    \centering
    \caption{Growth and characterization of epitaxial CeSi$_{2}$ films on Si(100).
    (a) Three-dimensional view of one u.c. CeSi$_{2}$, which consists of four Ce-Si-Si trilayers stacked along [001] with alternating in-plane positions.
    (b) The three-dimensional {bulk BZ (black) and the two-dimensional surface BZ (red). The surface BZ corresponds to the reciprocal lattice of the top layer and is often used for presenting the ARPES data.}
    (c) RHEED patterns of a 2x1 reconstructed Si(100) substrate and an as-grown CeSi$_{2}$ film.
    (d) XRD scans for the bare Si(100) substrate (black curve) and two representative thick CeSi$_{2}$ films (red curves) taken along $z$. The substrate peaks are labelled with black asterisks and the film peaks are labelled in red. Note that the X-ray photons contain a tiny contribution from Cu $K$$_{\beta}$ lines, in addition to the dominant Cu $K$$_{\alpha}$ line.
    (e) A STEM image of a CeSi$_{2}$ film near the interface, taken with the HAADF mode.
    (f) A high-resolution view of atomic structures in the dashed light blue box of (e), taken with the iDPC mode. The observed Ce and Si atoms show good agreement with the expected atomic structures in CeSi$_{2}$ and Si, as indicated on the left. {Here the in-plane direction is along the [120] direction of CeSi$_{2}$ and the [130] direction of Si. This indicates that the in-plane [100] direction of CeSi$_{2}$ aligns with the Si [110] direction, consistent with RHEED results.}
    (g) Resistivity of a thick CeSi$_{2}$ film as a function of temperature. Inset is a zoom-in view at very low temperature, indicating the absence of magnetic order.
    }
\label{Fig1}
\end{figure*}

\par In strongly correlated $f$-electron systems, the pronounced interactions between conduction electrons and localized $f$ electrons can lead to coherent quasiparticles with large effective mass at low temperatures, commonly called heavy fermions. Heavy-fermion metals are prototypical systems to study quantum phase transitions, quantum criticality and unconventional superconductivity \cite{Coleman2007heavy,gegenwart2008quantum,Sachdev2011Quantum,Mathur1998Magnetically,JPSJ2007review,Lohneysen2007Fermi,WHITE2015review,kirchner2020colloquium,Smidman2023RMP}. It was appreciated early on that a reduced dimensionality can be key to enhance the electron correlations and spin fluctuations in heavy fermion compounds \cite{coleman2010Science,Thompson2011NP}, thereby providing an important control knob to enhance the transition temperature of heavy fermion superconductors. A well-known example is the ``quasi-two-dimensional'' tetragonal heavy-fermion superconductor CeCoIn${_5}$ with $T_c$ = 2 K, which is boosted from the pressure-induced superconductivity at $T_c$ $\approx$ 0.3 K in the three-dimensional antiferromagnetic cubic metal CeIn${_3}$. However, except for the recently discovered compound CeSiI \cite{Posey2024}, most of the known heavy-fermion compounds, including CeCoIn${_5}$ and its Ce-115 homologues, are not really two-dimensional. Instead, they are anisotropic three-dimensional systems, which exhibit noticeable $k_z$ dispersion that must be taken into account. Therefore, to investigate the truly two-dimensional heavy-fermion physics, it is often necessary to grow ultrathin layers or superlattices to spatially confine the $f$ electrons. While there are only a few successful cases, i.e., CeIn${_3}$/LaIn${_3}$ \cite{Shishido2010Science}, CeCoIn${_5}$/YbCoIn${_5}$ \cite{Mizukami2011} and CeCoIn${_5}$/CeRhIn${_5}$ superlattices \cite{PhysRevLett.120.187002}, YbCu${_2}$ single layer \cite{Nakamura2023NC}, how the heavy-fermion state evolves from three to two dimensions remains elusive in electron spectroscopy. In addition, how the evolution of heavy-fermion states can be linked to changes in transport properties is still an open question.

\par Our study is also motivated by recent advances in two-dimensional heavy-fermion systems derived from artificial structures, most notably twisted bilayer graphene \cite{Cao2018second,PhysRevB.106.245129,PhysRevLett.129.047601}, stacked superlattices made from van der Waals materials \cite{Vano2021,Fan2024} {and terraced surfaces confined by steps \cite{Herrera2023}}. While heavy-fermion states in the former two cases are still attributed to Kondo coupling, the local moments and itinerant conduction electrons are more separated spatially compared to conventional heavy-fermion systems, possibly leading to different ramifications of the heavy quasiparticles. In addition, since these two-dimensional heavy-fermion systems are created from artificial structures, it is difficult to find their counterparts in three dimensions.

\par In this paper, we have studied the evolution of heavy quasiparticles in the prototypical heavy-fermion compound CeSi$_{2}$, continuously from three-dimensional thick films to two-dimensional ultrathin films. {Our study is made possible by a combination of MBE growth of high-quality ultrathin films, $in$-$situ$ ARPES measurements of the electronic states \cite{Chatterjee2017} and systematic transport measurements. While ARPES measurements reveal the one-electron spectral function in heavy fermion systems, which features a Kondo peak (or Abrikosov-Suhl resonance) at $E_F$ and CEF satellite peaks located both above and below $E_F$, complementary transport measurements are crucial to understand the manifestations of heavy-fermion states in physical properties. This combination is important for a comprehensive understanding of the physics.} We found that, from three to two dimensions, the ground-state Kondo peak remains strong at sufficiently low temperatures, indicating resilient quasiparticles in the two-dimensional limit. On the other hand, {the quantum confinement along the $z$ direction significantly influences the Kondo process at intermediate temperatures, where the involvement of CEF excitations becomes different and profoundly affects the transport properties.}

\section{RESULTS}
\subsection{Growth and characterization of CeSi$_{2}$ films}

\par Nearly stoichiometric CeSi$_{2}$ has a tetragonal crystal structure and is a paramagnetic heavy-fermion compound with an estimated single-ion Kondo temperature $T_K$ $\sim$ 40 K \cite{GALERA1989801}. One unit cell (u.c.) of a (001)-oriented CeSi$_{2}$ film consists of four Ce-Si-Si tri-layers in the $x$-$y$ plane (Fig. \ref{Fig1}(a)), with alternating in-plane stackings. Despite the challenge of synthesizing bulk single crystals of CeSi$_{2}$ \cite{SOUPTEL2004606,Shin2023arXiv}, we are able to grow epitaxial CeSi$_{2}$ films on Si(100) substrates using MBE. The structural quality of the CeSi$_{2}$ films can be verified by the sharp patterns from reflection high-energy electron diffraction (RHEED) taken during the film growth, as shown in Fig. \ref{Fig1}(c). RHEED patterns taken at different thicknesses further verify that the films remain of good quality from thin to thick films (see {Fig. S1 in \cite{supplementary}}). Figure \ref{Fig1}(d) shows the X-ray diffraction (XRD) scans of thick CeSi$_{2}$ films, which exhibit sharp Bragg peaks from CeSi$_{2}$ with an lattice constant $c$ of 13.9 {\AA}. The film quality can be best visualized by measurements from scanning transmission electron microscopy (STEM) using the high-angle annular dark-field scanning mode (HAADF) (Fig. \ref{Fig1}(e)), where atomically smooth films over a large scale can be directly observed. Fig. \ref{Fig1}(f) displays a zoom-in view of the atomic images near the interface using the integrated differential phase contrast (iDPC) mode, with both Ce and Si atoms near the interface being clearly identified. The atomic positions are in good agreement with the expected CeSi$_{2}$ and Si lattices (see atomic models on the left of Fig. \ref{Fig1}(f)). A detailed analysis of the STEM images (see {Fig. S2 in \cite{supplementary}}) indicates that the lattice constant of the interfacial CeSi$_{2}$ layer is already close to the bulk value, with $a$ $\approx$ 4.2 {\AA} and $c$ $\approx$ 13.9 {\AA}. This implies that even for 1 u.c. film of CeSi$_{2}$, the crystal structure is already relaxed (or close to bulk), probably due to the large lattice mismatch between CeSi$_{2}$ and the Si(100) substrate ($\sim10\%$).

\par Figure \ref{Fig1}(g) displays the temperature-dependent resistivity for a typical thick film of CeSi$_{2}$, which indicates a clear metallic behavior with no abrupt transition down to the lowest measurement temperature 0.3 K (see inset of Fig. \ref{Fig1}(g)). This is consistent with the results of nearly stoichiometric CeSi$_{2}$ with a paramagnetic ground state \cite{P-Hill1992JPCM} - a small Si deficiency would otherwise lead to ferromagnetic ordering at very low temperatures \cite{P-Hill1992JPCM,PhysRevB.35.6880,SOUPTEL2004606,Shin2023arXiv}. It is worth mentioning that Si deficiency is common when synthesizing bulk CeSi$_{2}$ crystals. This is circumvented here, since the Si substrate always provides Si-rich growth conditions.

\begin{figure*}[ht]
    \includegraphics[scale=0.75]{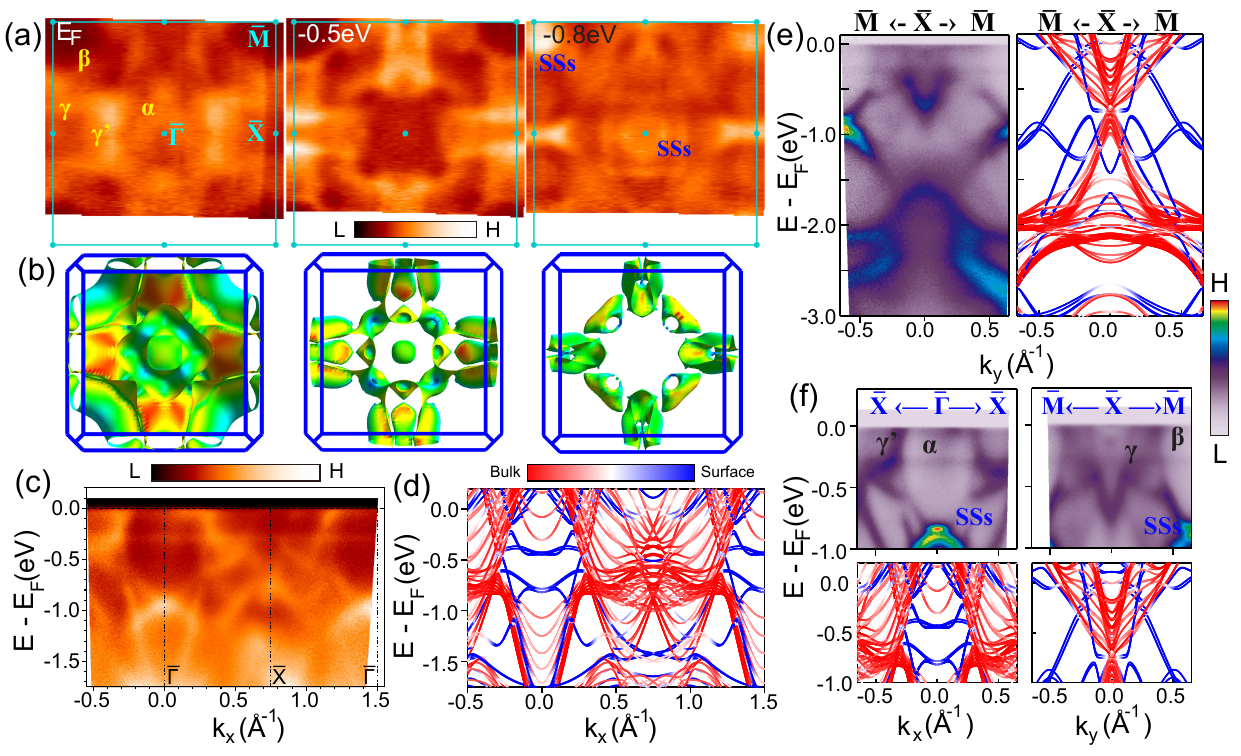}
    \centering
    \caption{Fermi surface and valence bands of thick CeSi$_{2}$ films and comparison with DFT calculations.
    (a) Constant-energy $k_x$-$k_y$ maps at $E$ = $E_F$ (0 eV), -0.5 eV and -0.8 eV. The surface BZ is indicated by light green box. The bulk bands ($\alpha$, $\beta$, $\gamma$ and $\gamma$') and surface states (SSs) are also labelled.
    {(b) Constant-energy $k_x$-$k_y$ contours from DFT calculations of bulk CeSi$_{2}$ for comparison with (a). Contour colors indicate different Fermi velocities. }
    (c) Large-range energy-momentum dispersion along $\bar{\Gamma}-\bar{X}-\bar{\Gamma}$.
    (d) DFT slab calculations, for comparison with (c).
    (e) Energy-momentum dispersion along $\bar{M}-\bar{X}-\bar{M}$ (left), in comparison with slab calculation (right).
    (f) Zoom-in view of band disperison (top panels) near $E_F$ along $\bar{X}-\bar{\Gamma}-\bar{X}$ and $\bar{M}-\bar{X}-\bar{M}$, in comparison with slab calculations (bottom panels).
    All experimental data were taken with 21.2 eV photons. Curve colors in (d-f) indicate surface (blue) or bulk (red) contributions. {The calculated bulk bands in (b,d-f) include contributions from all $k_z$'s.}
    \label{Fig2}
    }
\end{figure*}

\subsection{Fermi surface and valence bands of thick films}

\par Our synthesis of epitaxial CeSi$_{2}$ films with flat surfaces allows for direct measurements of the momentum-resolved electronic structure by ARPES. Figure \ref{Fig2} shows the ARPES spectra of a typical thick CeSi$_{2}$ film taken with 21.2 eV (He I) photons; {the corresponding surface Brillouin zone (BZ) is displayed accordingly in Fig. \ref{Fig1}(b).} In Fig. \ref{Fig2}(a), the evolution of the constant-energy $k_x$-$k_y$ maps is displayed at a few representative energies. Here the measured Fermi surface (FS, $E$ = 0) consists of a small pocket centered at $\bar{\Gamma}$ (labelled $\alpha$), another circular pocket centered at $\bar{M}$ (labelled $\beta$) and two sets of bands running parallel to $k_x$ and $k_y$ axes (labelled $\gamma$ and $\gamma$'), respectively. {The measured FS is in good agreement with the bulk Fermi contour from density functional theory (DFT) calculations, as shown in Fig. \ref{Fig2}(b), where the projection from all $k_z$'s is included (due to possible large $k_z$ broadening in He I ARPES measurements).} However, for the maps away from the Fermi energy ($E_F$), some experimental features cannot be explained by the calculations of the bulk bands alone, e.g., the rich structures observed near the $\bar{\Gamma}$ and $\bar{M}$ points at $E$ = -0.8 eV. These features can be attributed to surface states (SSs), as we shall discuss below.

\par Figures \ref{Fig2}(c,e) show the ARPES spectra over a large energy range taken along the $\bar{\Gamma}-\bar{X}-\bar{\Gamma}$ and $\bar{M}-\bar{X}-\bar{M}$ directions, respectively. The corresponding zoom-in views near $E_F$ are displayed in Fig. \ref{Fig2}(f). Note that the flat bands near $E_F$ and -0.3 eV are derived from Ce 4$f$ electrons, which we shall discuss in detail later. To understand the bulk valence bands and distinguish SSs from bulk bands, we have also performed slab calculations using DFT, where a thick film with vacuum-terminated surfaces was used to calculate both SSs and bulk bands (including all $k_z$'s). The results are displayed in Figs. \ref{Fig2}(d-f) for direct comparison with the experimental data (details included in {Fig. S3 in \cite{supplementary}}). Here most of the valence bands observed near $E_F$ show good agreement with the bulk bands from slab calculations (red curves in Fig. \ref{Fig2}(f)). This naturally explains why the experimental FS in Fig. \ref{Fig2}(a) can be well captured by bulk DFT calculations alone (Fig. \ref{Fig2}(b)).On the other hand, there are a few sharp bands away from $E_F$ that cannot be explained by bulk bands, e.g., the hole band near -0.9 eV at $\bar{\Gamma}$ (Fig. \ref{Fig2}(c,d)) and the sharp bands near $\sim$-1 eV at $\bar{M}$ (Fig. \ref{Fig2}(e)) - these bands can be attributed to SSs, based on comparison with slab calculations (blue curves in Fig. \ref{Fig2}(d-f)). {Since these SSs, mostly derived from Si 3$p$ and Ce 5$d$ orbitals (see Fig. S4 in \cite{supplementary}), are far from the 4$f$ bands near $E_F$, they do not hybridize with the 4$f$ electrons and hence show weak correlation effects.} We note that the calculated SSs are dependent on the surface terminations (see Fig. S3 in \cite{supplementary}). The observed SSs can only be explained by a Si-terminated surface with a Si adlayer on top, which is often found for transition-metal or rare-earth silicide films grown on Si substrates by MBE \cite{PhysRevB.37.10786,PhysRevB.61.5707,WOOD2005120,Fang_2021}. The Si-terminated surface and the bulk-band-dominated FS ensure that the measured $4f$ states near $E_F$ reflect the bulk property of CeSi$_{2}$.

\begin{figure*}[ht]
    \includegraphics[width=1.7\columnwidth]{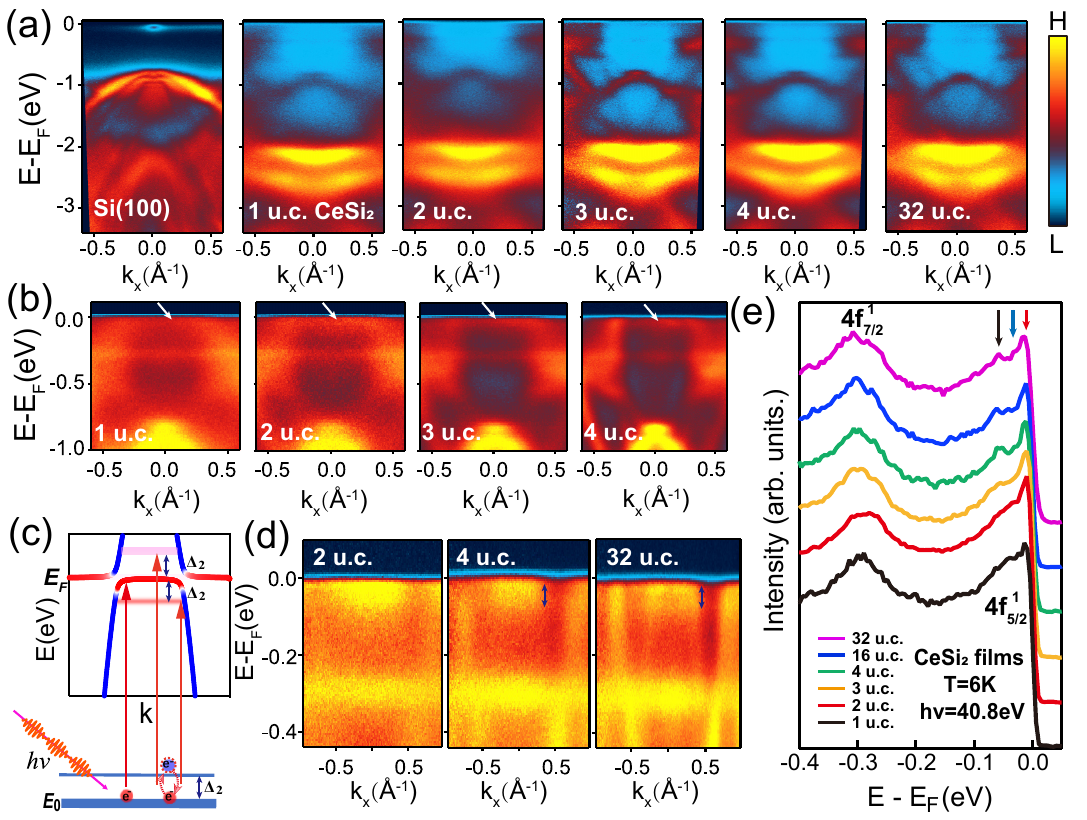}
    \centering
    \caption{The thickness-dependent electronic structure of CeSi$_{2}$ films along $\bar{X}-\bar{\Gamma}-\bar{X}$ at 6 K.
    (a) Thickness evolution of the valence bands taken with 21.2 eV photons. The valence bands from the Si(100) substrate are also shown for comparison.
    (b) Zoom-in view near $E_F$ from (a), from 1 u.c. to 4 u.c.. White arrows indicate the reversed U-shaped bands from hybridization between $4f$ and valence bands.
    (c) Top panel: a schematic plot of the quasiparticle dispersion as a result of hybridization between $4f$ (red curve) and valence (blue curve) bands in a periodic Anderson model. The pink and red diffuse lines indicate the CEF satellites above and below $E_F$. Bottom panel: a schematic diagram illustrating the origin of the CEF satellites.
    (d) Thickness evolution of the quasiparticle bands near $E_F$ taken with 40.8 eV photons, highlighting the $4f$ contributions.
    (e) The EDCs near $\bar{\Gamma}$ as a function of thickness, taken with 40.8 eV photons. The EDCs are integrated from -0.25 {\AA}$^{-1}$ to 0.25 {\AA}$^{-1}$. The ground-state Kondo peak and CEF satellites are labelled by red and black/blue arrows, respectively. The EDCs are shifted vertically for different thicknesses.
    }
\label{Fig3}
\end{figure*}

\subsection{Electronic structure evolution with film thickness}

\par Figure \ref{Fig3}(a) shows the evolution of the valence bands as a function of film thickness, taken with 21.2 eV photons. Upon deposition of the 1 u.c. CeSi$_{2}$ film, the valence bands of the Si substrate completely disappear and the characteristic valence bands of CeSi$_{2}$ are formed. The fact that no trace of valence bands from the Si substrate can be observed at 1 u.c. implies that the 1 u.c. CeSi$_{2}$ film is smooth and mostly covers the Si(100) substrate. With further deposition of CeSi$_{2}$, the SS at $\sim$-1 eV remains similar, while the bulk bands between -2 eV and -3 eV become broader. Figure \ref{Fig3}(b) displays zoom-in views of the hole-type valence bands near $E_F$ (additional data in Fig. S5 in \cite{supplementary}), which are mostly derived from (in-plane) Ce $d_{xy}$ and Si $p_{x}$/$p_{y}$ orbitals, based on orbital analysis from DFT. {From detailed photon-energy-dependent study (Fig. S6 in \cite{supplementary}), these valence bands exhibit clear $k_z$ dispersion for thick films but much weaker $k_z$ dependence for thin films, confirming the thickness-driven dimensionality crossover of the electronic states.} Note that quasiparticle bands can be well observed down to 1 u.c. - this demonstrates the high quality of ultrathin CeSi$_{2}$ films and allows us to investigate the intrinsic effects resulting from dimensionality tuning.

\par From Fig. \ref{Fig3}(b), we can already identify two sets of (flat) $4f$ bands near $E_F$, one at $E_F$ ($4f^1_{5/2}$) and another at $\sim$-0.3 eV ($4f^1_{7/2}$), which originate from the many-body Kondo process including the spin-orbit coupling (SOC) of the Ce $4f$ electrons \cite{PhysRevLett.58.2810,jw2005kondo,fujimori2016band}. For a periodic Kondo lattice at sufficiently low temperatures, the ground-state $4f$ band near $E_F$ should develop momentum-dependent dispersion due to hybridization with valence bands \cite{denlinger2001comparative,PhysRevX.5.011028,patil2016arpes,Chen2017,Chen2018,Jang2020,PhysRevX.14.021048}, as illustrated in Fig. \ref{Fig3}(c) (top panel). Such a momentum dependence can be readily identified from the experimental data in Fig. \ref{Fig3}(b): a hole-type valence band ($\gamma$') hybridizes with the $4f^1_{5/2}$ band at $E_F$, leading to a reversed U-shaped band centered at $\bar{\Gamma}$ (marked by white arrows in Fig. \ref{Fig3}(b)).

\par
\begin{figure*}[ht]
    \includegraphics[width=2.0\columnwidth]{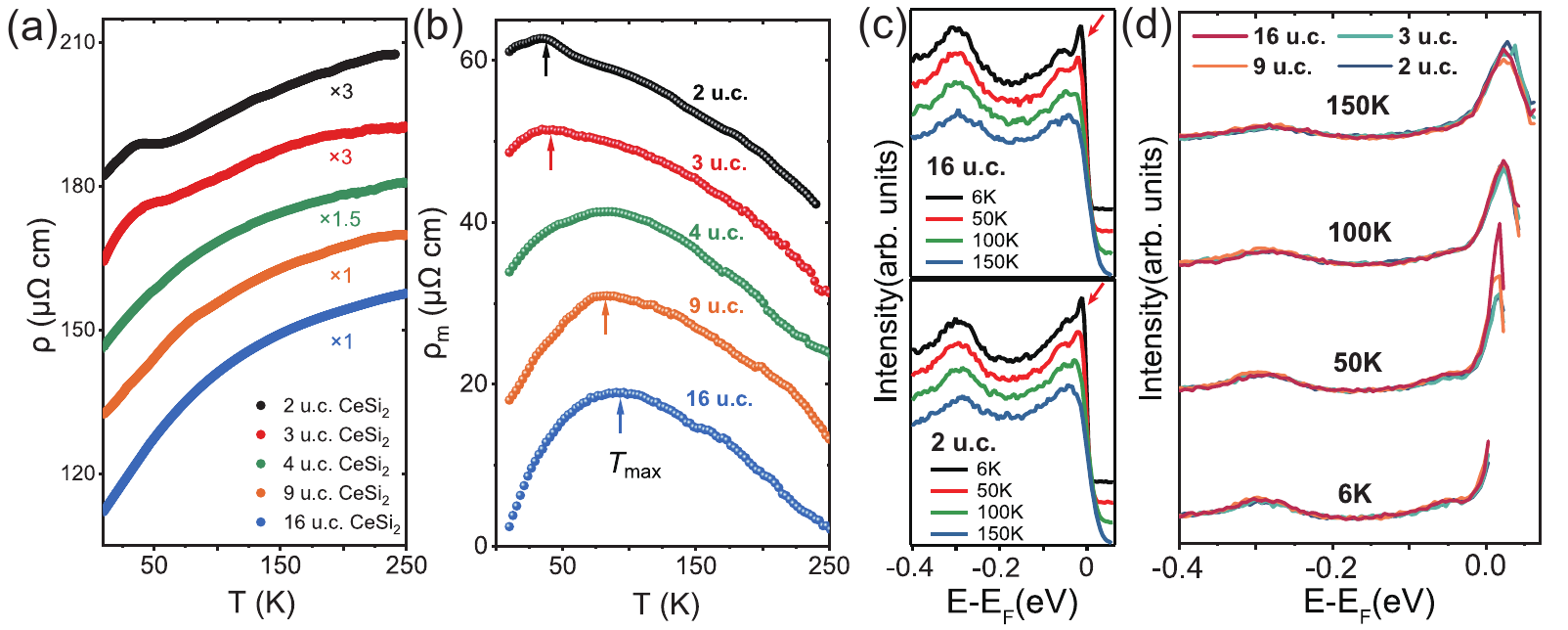}
    \centering
    \caption{Temperature evolution of the resistivity and $4f$ states.
    {(a) The temperature-dependent resistivity for CeSi$_{2}$ films with different thicknesses. The absolute values of resistivity were multiplied by a prefactor shown for each curve, so that the data can be displayed properly in one plot.
    (b) The extracted magnetic part of resistivity $\rho_m$, from the data in (a). $\rho_m$ is obtained by subtracting the resistivity of LaSi$_{2}$ films from that of CeSi$_{2}$ films with identical thickness. Arrows indicate the positions of the maximum $T_{max}$.}
    (c) The temperature-dependent EDCs near $\bar{\Gamma}$ for 16 u.c. (top) and 2 u.c. (bottom) films, taken with 40.8 eV photons. The EDCs are integrated from -0.25 {\AA}$^{-1}$ to 0.25 {\AA}$^{-1}$.
    (d) The EDCs from (c) divided by the RC-FDD function, for different films at 6 K, 50 K and 100 K, respectively. Only the energy region of $E <$ 6$k_B$$T$ is shown here.
    All data are offset vertically for display clarity.
    }
\label{Fig4}
\end{figure*}

 Due to the small photoemission cross section of $4f$ electrons under He I photons, we also employ He II photons (40.8 eV) to perform ARPES measurements, where the spectral contributions from the $4f$ electrons are much enhanced. The quasiparticle dispersions obtained from He II photons are summarized in Fig. \ref{Fig3}(d) for a few representative thicknesses. For the 32 u.c. and 4 u.c. films, the $4f^1_{5/2}$ band near $E_F$ appears to split into two bands (highlighted by blue arrows in Fig. \ref{Fig3}(d)). For the 2 u.c. films, the splitting becomes difficult to identify. For a detailed analysis, we plot in Fig. \ref{Fig3}(e) the energy distribution curves (EDCs) near $\bar{\Gamma}$ as a function of film thickness. For thick films, e.g., 4 u. c. and above, the $4f^1_{5/2}$ peak indeed consists of a well-defined peak near $E_F$ (red arrow) and another satellite peak at $\sim$-55 meV (black arrow). While the peak near $E_F$ corresponds to the ground-state Kondo peak that is situated slightly above $E_F$ and is truncated by the Fermi-Dirac function, the satellite peak at $\sim$-55 meV can be attributed to the (virtual) transition between the ground-state and an excited CEF doublet \cite{patil2016arpes,KROHA200369,ehm2007high,PhysRevB.97.075149}, as illustrated in Fig. \ref{Fig3}(c) (bottom panel): In the Kondo spin fluctuation processes, the (virtual) transition from the ground-state to the excited CEF doublet can lead to a CEF satellite above $E_F$ (+$\Delta_2$), and the reverse transition (from the excited CEF doublet back to the ground-state doublet) results in a CEF satellite lying below $E_F$ (-$\Delta_2$) \cite{KROHA200369,PhysRevLett.122.096401}. Although these transitions are suppressed thermally at low temperature (6 K in Fig. \ref{Fig3}(e)), they are still possible due to the quantum mechanical nature of Kondo spin fluctuations \cite{KROHA200369,PhysRevLett.122.096401}. {Indeed, CEF satellites below $E_F$ have been observed at low temperatures in CeRh$_{2}$Si$_{2}$ \cite{patil2016arpes} and CeIrIn$_{5}$ \cite{PhysRevB.97.075149}, and they are also observed in CeSi$_{2}$ films here (Fig. \ref{Fig3}(d)). While the CEF satellite at $\sim$-55 meV is weak at 6 K (albeit fully accessible by ARPES), the CEF satellite above $E_F$ can only be seen by ARPES at elevated temperatures (see Fig. \ref{Fig4}(d) for detailed discussions).}

\par The position of the satellite peak ($\sim$-55 meV) is consistent with the CEF splitting of CeSi$_{2}$ obtained from inelastic neutron scattering (INS) \cite{GALERA1989801,ehm2007high}: for tetragonal CeSi$_{2}$, the CEF scheme of the Ce$^{3+}$ $4f$ electrons consists of a ground-state doublet and a first (second) excited doublet with an estimated energy separation $\Delta_1$ $\sim$25 meV ($\Delta_2$ $\sim$50 meV) from the ground state \cite{GALERA1989801}. Therefore, the observed peak separation is in reasonable agreement with $\Delta_2$ (considering the ARPES energy resolution and uncertainties in the INS fittings \cite{GALERA1989801}), while a very weak shoulder seems to be present near $\sim$-30 meV (blue arrow in Fig. \ref{Fig3}(e), {see also Fig. S7 in \cite{supplementary}}) and it might be caused by the excitation to the first excited doublet. Note that these fine structures below $E_F$ have not been resolved before in polycrystal samples \cite{PhysRevLett.58.2810,ehm2007high,PhysRevB.47.15460}.

\par As the film thickness decreases to 3 and 2 u.c., the ground-state Kondo peak at $E_F$ remains strong. However, the -55 meV satellite becomes largely suppressed below 3 u.c. (see Fig. \ref{Fig3}(e)), with possibly a broad shoulder remaining at the same energy. {This implies that the thickness reduction suppresses the Kondo effect involving the second excited doublet, although the energy separation of (local) CEF levels are likely unchanged. The robust ground-state Kondo peak at 2-3 u.c. rules out the possibility that the suppressed CEF satellite here originates from disorder, as disorder would lead to similar suppression of the ground-state Kondo peak. Our ARPES and STEM measurements also reveal similar sample quality from thick films down to 3 u.c. (see Fig. \ref{Fig3}(a,b) and Fig. S2(a) in \cite{supplementary}). On the other hand, the ground-state Kondo peak for the 1 u.c. film becomes clearly weaker compared to thick films, and its suppression is likely related to the disorder effect: although all CeSi$_{2}$ films were grown under identical conditions, interfacial disorder can be present due to interfacial defects and should be largely limited to the interfacial layer. Therefore, despite the possible disorder effect for the 1 u.c. film, our systematic thickness-dependent data demonstrate that the many-body Kondo process associated with the ground-state doublet remains robust from three to two dimensions, but the Kondo process involving CEF excitations becomes obviously suppressed (see below for more discussions). }

\subsection{Temperature evolution of the resistivity and $4f$ states}

\par Figure \ref{Fig4}(a) displays the temperature-dependent resistivity for CeSi$_{2}$ films with different thicknesses (additional data included in {Fig. S8 in \cite{supplementary}}). {For comparison, reference LaSi$_{2}$ films with the same thicknesses are grown and measured under identical conditions (Fig. S8 in \cite{supplementary}).} While the resistivity of LaSi$_{2}$ films shows typical metallic behavior with dominant contributions from phonon scattering \cite{TRAVLOS199713,PhysRevB.100.165308}, the resistivity of thick CeSi$_{2}$ films exhibits a slope change (or broad crossover) slightly below 100 K, similar to bulk CeSi$_{2}$ \cite{SOUPTEL2004606}. When the film thickness decreases to 3 or 2 u.c., the crossover becomes sharper and its temperature decreases to $\sim$35 K. To extract the magnetic part of the resistivity, $\rho_m$($T$), {we subtract the resistivity of LaSi$_{2}$ films from that of CeSi$_{2}$ films (see methods for details)} and the results are summarized in Fig. \ref{Fig4}(b). It is clear that $\rho_m$($T$) for all CeSi$_{2}$ films exhibits a broad peak, defined as $T_{max}$ (see arrows in Fig. \ref{Fig4}(b)), which is a hallmark of heavy-fermion metals. We mention that the resistivity of 1 u.c. film is not included here, {since it can be affected by the percolation issue that becomes important in 1 u.c. (see Fig. S8 and S9 in \cite{supplementary}).}

\par For a given film thickness, $\rho_m$($T$) is found to rise upon cooling at temperatures $T > T_{max}$, due to incoherent Kondo scattering, while it decreases below $T_{max}$. For the films thicker than 4 u.c., this decrease below $T_{max}$ is not simply from the lattice coherence, as demonstrated, e.g., in CeCoIn${_5}$ \cite{Chen2017}. Instead, it mainly reflects the thermal depopulation of excited CEF states \cite{PhysRevB.5.4541,RevModPhys.56.755,PhysRevB.77.104412}, and merges with contributions from the lattice coherence at lower temperatures. Such a peak in $\rho_m$($T$) often resides at $T_{max}$ = 0.3-0.6 $\Delta_{CEF}$/$k_B$ \cite{PhysRevB.77.104412}, which is overall consistent with $T_{max}$ $\approx$ 100 K for $\Delta_1$ $\sim$ 25 meV here. A decrease of $T_{max}$ from $\approx$100 K to $\approx$35 K is observed in Fig. \ref{Fig4}(b) upon decreasing the film thickness, which can be attributed to the dimensionality tuning of the Kondo process, i.e., a reduction of Kondo scattering channels along the $z$ direction: this causes a reduction of the ``effective'' Kondo energy from the four-fold or six-fold $4f$ states (including excited CEF states) in case of the three-dimensional thick films to the value of $k_B$$T_k$ with degeneracy $N = 2$ in case of the two-dimensional ultrathin films (see below for more discussions). Indeed, $T_{max}$ $\approx$ 35 K for 3 u.c. and 2 u.c. films agrees reasonably well with the single-ion Kondo temperature $T_K$ $\sim$ 40 K (corresponding to the ground-state doublet) obtained from neutron scattering \cite{GALERA1989801}. Therefore, the ultrathin films can be regarded as truly two-dimensional heavy-fermion systems - with in-plane transport involving conduction-electron scattering mostly from the 4f-electron ground-state doublet. The thickness dependence of $T_{max}$ also corresponds well to the presence (absence) of CEF satellites below $E_F$ in thick (thin) films, as shown in Fig. \ref{Fig3}(e), implying that the difference in the CEF excitations is concommitant with the dimensionality transition.

%When CEF excitations are absent or weak, the peak at $T_{max}$ mostly reflects the onset of lattice coherence: $\rho_m$($T$) rises upon cooling at temperature $T$ > $T_{max}$, due to incoherent Kondo scattering, while it decreases below $T_{max}$, which is mostly caused by lattice coherence in Kondo scattering. This applies to the 2 u.c. and 3 u.c. films, where $T_{max}$ $\sim$ 35 K agrees reasonably well with the single-ion Kondo temperature $T_K$ $\sim$ 40 K obtained from neutron scattering \cite{GALERA1989801,PhysRevB.35.6880}. On the other hand, if CEF excitations are strongly involved in the Kondo process, thermal depopulation of excited CEF states upon cooling can lead to reduced scatterings already at higher temperatures \cite{PhysRevB.5.4541,RevModPhys.56.755,PhysRevB.77.104412}: if the crossover is broad, this can result in a single peak in $\rho_m$($T$) with increased $T_{max}$ compared to $T_K$.  In the current case, the obvious increase of $T_{max}$ in thicker CeSi$_{2}$ films can be similarly attributed to scatterings from excited CEF states, while a weak slope change at 35 K (highlighted by triangles in Fig. \ref{Fig4}(b)) remains visible, corresponding to coherence.

\par Figure \ref{Fig4}(c) displays the temperature-dependent photoemission data for 16 u.c. and 2 u.c. films (see Fig. S10 in \cite{supplementary} for additional data). For the 16 u.c. film, the ground-state Kondo peak at $E_F$ (marked by a red arrow in Fig. \ref{Fig4}(c)) becomes weaker and broadened with increasing temperature, consistent with its many-body nature. As expected, this Kondo peak merges with the satellite at $\sim$-55 meV at elevated temperatures. A similar temperature evolution can be observed for the 2 u.c. film, despite a much weaker shoulder at $\sim$-55 meV. To recover the full spectral function near $E_F$, we divided the experimental EDCs by the Resolution-Convoluted Fermi-Dirac Distribution (RC-FDD) function, determined by measuring an Au reference sample (see {Fig. S10 in \cite{supplementary}}). The recovered spectral functions for representative thicknesses are summarized in Fig. \ref{Fig4}(d). At 100 K, the Kondo process and thermal population of the excited CEF states (mainly $\Delta_1$) lead to a broad Kondo peak above $E_F$ \cite{ehm2007high,Chen2017}, although it is difficult to resolve each CEF satellite here due to the large energy broadening. This observation is consistent with the $\rho_m$($T$) results displayed in Fig. \ref{Fig4}(b), which indicate mostly incoherent Kondo scatterings at high temperatures. Note that the broad Kondo peak here still contains a sizable contribution from the ground-state doublet, and hence its variation with thickness is not as obvious as the individual CEF satellite observed below $E_F$.

\par As already mentioned, upon cooling across $T_{max}$, the excited CEF states become thermally depopulated. Partial lattice coherence is also anticipated to set in over a broad temperature range, as observed by ARPES measurements on the canonical heavy-fermion metals CeCoIn$_{5}$ \cite{Chen2017,Jang2020} and CeRhIn$_{5}$ \cite{Chen2018}. An interesting observation is made in Fig. \ref{Fig4}(d) at an intermediate temperature of 50 K: the Kondo peak near $E_F$ is obviously stronger for thicker films. This implies that the aforementioned partial lattice coherence (or the "coherent quasiparticle") is developed the better, the thicker the film is. This difference in the Kondo peak intensity at 50 K correlates well with the large thickness dependence of $\rho_m$($T$ = 50 K), suggesting an effective enhancement of the Kondo process for thicker films due to contributions from excited CEF states. On further cooling to temperatures well below $T_{max}$, a coherent Kondo peak derived from the ground-state doublet should be formed at $\sim$$k_B$$T_K$ above $E_F$ \cite{Ernst2011,PhysRevLett.108.066405}. Experimentally, because of the Fermi-Dirac cutoff, the recovered ARPES spectral function at 6 K can only access part of the full Kondo peak (see Fig. \ref{Fig4}(d)). Therefore, it is difficult to make a reliable comparison between different thicknesses at 6 K.

\section{Discussion}

\begin{figure*}[ht]
    \includegraphics[width=1.5\columnwidth]{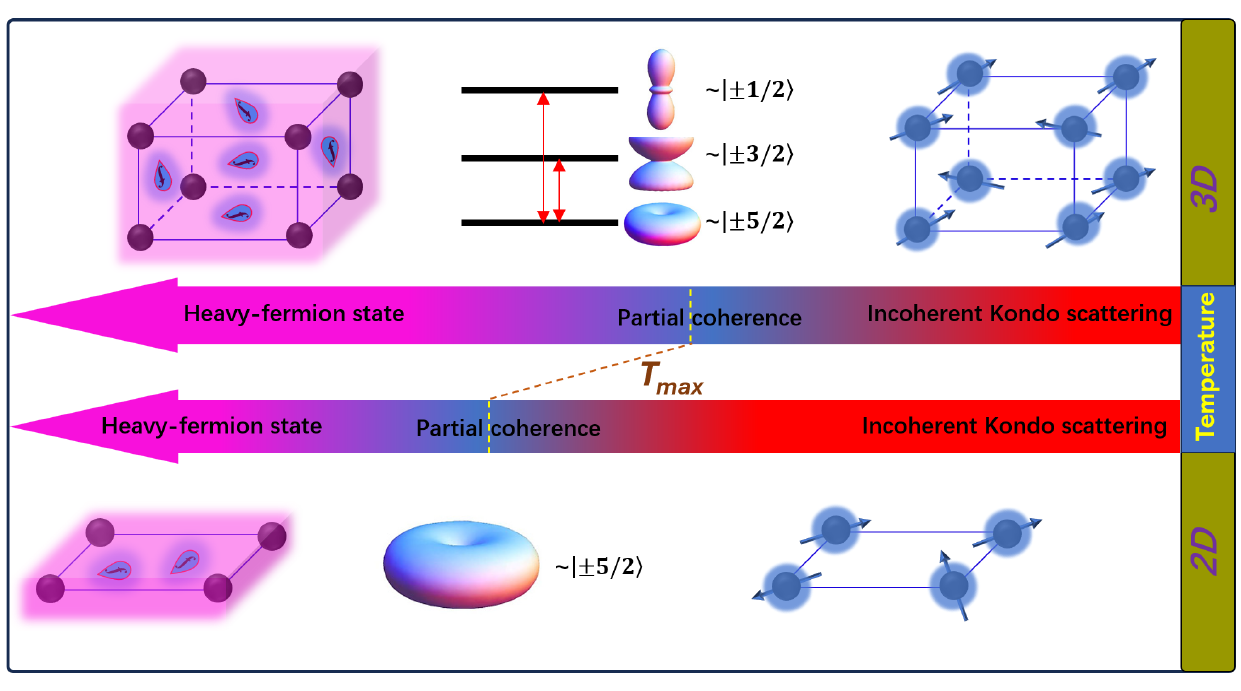}
    \centering
    \caption{Schematic diagrams illustrating the dimensionality tuning of the temperature-dependent Kondo process in CeSi$_{2}$.
    From three to two dimensions, {the many-body Kondo effect at intermediate temperatures becomes diminished due to reduced participations of excited CEF states (see middle panels)}, resulting in a decrease of $T_{max}$ in $\rho_m$($T$). Nevertheless, the heavy-fermion state at low temperatures, derived from the $|\pm5/2>$ ground-state doublet, remains robust in the two-dimensional limit.
    }
\label{Fig5}
\end{figure*}

\par The observed thickness dependence of the $4f$ states and the transport properties (Fig. \ref{Fig3} and Fig. \ref{Fig4}) can be attributed to the dimensionality tuning of the Kondo effect - see Fig. \ref{Fig5} for a schematic illustration. Note that the lattice constant of the 1 u.c. film is almost identical to that of thick films (see {Fig. S1 and S2 in \cite{supplementary}}). Therefore, the dimensional transition does not involve any complication from the change in lattice constants. In the two-dimensional limit, due to the diminishing partners along the $z$ direction, the many-body Kondo effect may be thought to be suppressed due to the decreasing number of conduction electrons (along $z$) available for the Kondo screening. Yet, the resilient heavy-fermion state observed in ultrathin CeSi$_{2}$ films indicates that two-dimensional conduction electrons within the $x-y$ plane are sufficient to screen the local moments, although the development of the heavy quasiparticles might be pushed to lower temperatures in the two-dimensional limit (see Fig. \ref{Fig4}(d)).

\par In addition to heavy-fermion thin films grown by MBE \cite{Shishido2010Science,Mizukami2011,PhysRevLett.120.187002}, two-dimensional heavy fermions have been recently reported in the layered van der Waals compound CeSiI \cite{Posey2024} and other interfacial two-dimensional systems \cite{Cao2018second,PhysRevB.106.245129,PhysRevLett.129.047601,Vano2021,Fan2024,Herrera2023}, representing new frontiers in heavy-fermion research. While our study demonstrates that stronger electron correlations in two-dimensional heavy-fermion metals manifest through a reduced "effective" Kondo energy due to the suppression of CEF excitations, quantum fluctuations are also thought to be enhanced in two dimension, as verified in antiferromagnetic systems \cite{Shishido2010Science}. While our ARPES measurements are currently limited to 6 K, future studies at very low temperatures, e.g., from scanning tunneling microscopy (STM) and magneto-transport measurements, should help to explore possible quantum critical phenomena in these ultrathin CeSi$_{2}$ films. In particular, since stoichiometric bulk CeSi$_{2}$ is not far from a ferromagnetic quantum critical point \cite{Shin2023arXiv}, dimensionality tuning in thin films could provide a clean and alternative pathway to search for ferromagnetic quantum criticality, which is currently of considerable interests \cite{RevModPhys.88.025006,steppke2013science,Shen2020,PhysRevLett.134.126401}.

\par It is useful to analyze the wave function of the CEF states for a deeper understanding of two-dimensional heavy fermions in CeSi$_{2}$. Specifically, previous INS experiments have revealed that the ground-state doublet in CeSi$_{2}$ is mostly of $|m_{J}>$ = $|\pm5/2>$ character, while the first (second) excited doublet is mostly made of $|\pm3/2>$ ($|\pm1/2>$) \cite{GALERA1989801}, respectively. Since the wave function of the $|\pm5/2>$ state is distributed mostly within the $x-y$ plane (see Fig. \ref{Fig5}), it is natural to think that the lattice coherence involving this ground-state doublet can still take place in the two-dimensional limit, due to largely unaffected in-plane ``hoppings''. In this sense, the CEF scheme of CeSi$_{2}$ might be favorable for the formation of two-dimensional heavy fermions in ultrathin (001) films. It would be interesting to explore in the future the dimensionality tuning of heavy-fermion systems with different ground-state wave functions.

\par The thickness evolution of the satellite peak at $\sim$-55 meV is noteworthy (Fig. \ref{Fig3}(e)): it implies that the associated CEF excitation exhibits an abrupt change near 3 u.c. (or $\sim$4 nm). Note that the wave function of this excited CEF state (mostly $|\pm1/2>$) is largely directed along the $z$ direction (see Fig. \ref{Fig5}), {and therefore the Kondo process involving this CEF excitation might be more sensitive to the reduction of the film thickness}, compared to the ground-state Kondo process. Nevertheless, further in-depth studies are needed to understand this thickness dependence. In addition, future theoretical and experimental works are necessary to investigate how the development of partial lattice coherence and the augmentation of excited CEF states are interlinked as a function of film thickness.

\par {Finally, it is interesting to compare the dimensionality tuning of $4f$-electron heavy fermions with those derived from $5f$ electrons, where the correlation effects are often weaker. Recently, very fine quantum well states (QWSs) from $5f$ electrons have been observed from STM measurements on the stepped surfaces of URu$_{2}$Si$_{2}$ \cite{Herrera2023}, demonstrating the constructive interference of two-dimensional heavy quasiparticles with an effective mass of $\sim$17 $m_{e}$. Interestingly, these $5f$ quasiparticles can be well described by the Fabry-P\'erot interferometer with low reflection coefficient due to large effective mass \cite{Herrera2023}. In the current case of CeSi$_{2}$, the much larger effective mass of Ce $4f$ electrons ($\sim$150 $m_{e}$ estimated from the specific heat \cite{Shin2023arXiv}) and limited energy resolution of ARPES measurements make it extremely difficult to identify discrete QWSs from $4f$ electrons. Future high-resolution (STM) study will be desirable to seek for $4f$-electron QWSs.}

\section{CONCLUSION}

\par In summary, by combining MBE growth, ARPES and transport measurements, we have demonstrated the dimensionality tuning of the heavy-fermion state in the canonical heavy-fermion compound CeSi$_{2}$, from three all the way to two dimensions. Our work was made possible by the synthesis of high-quality ultrathin CeSi$_{2}$ films, where disorder effects can be minimized. Our results show that the Kondo process in the intermediate temperatures operates quite differently for the three-dimensional and two-dimensional cases, owing to the different involvement of excited CEF states. The ground-state Kondo peak at $E_F$ persists down to the two-dimensional limit, although its development is pushed towards lower temperatures. {Our work therefore highlights the importance of CEF excitations in the Kondo process for understanding low-dimensional heavy fermions. While most known heavy fermion compounds are anisotropic three-dimensional systems (including bulk CeSi$_{2}$), their truly two-dimensional counterparts are still rare with plenty of room for new discoveries. Our epitaxial CeSi$_{2}$ films with controllable thickness provide an ideal platform to study the dimensionality tuning of the heavy-fermion physics and explore the ferromagnetic quantum criticality in two dimension.}

\section{Methods}
\subsection{MBE growth and $in$-$situ$ ARPES measurements}
\par The epitaxial CeSi$_{2}$ films were grown by MBE on Si(100) substrates, modified from a previous growth recipe \cite{AOKI19921905}. The base pressure of the MBE growth chamber is better than $1\times10^{-10}$ mbar. The evaporation of Ce was achieved using effusion cells, whose deposition rates were set to approximately $\sim$1.5 {\AA}/min for Ce, as determined by a quartz crystal monitor (QCM). Before growth, Si(100) substrates were first heated to over 600 ℃ in ultrahigh vacuum for 30 minutes for degassing and then shortly flashed to over 1200 ℃ in a few seconds by passing through a direct current, leading to a sharp 2x1 RHEED pattern. Ce was then deposited on the 2x1 reconstructed Si(100) surface at $\sim$100 ℃ and further annealed to $\sim$600 ℃, allowing the Si from the substrate to react with the deposited Ce and form epitaxial CeSi$_{2}$ films. The film quality was monitored in real time by RHEED. The thickness of the CeSi$_{2}$ films was determined by the amount of Ce deposited on the subtrate and confirmed by subsequent STEM measurements.

\par After growth, the CeSi$_{2}$ films with different thicknesses were transferred under ultrahigh vacuum to an ARPES system at Zhejiang University for $in$-$situ$ ARPES measurements \cite{Zheng2025SCPMA}. The sample temperature was kept at $\sim6$ K during all measurements unless noted otherwise. The base pressure of the ARPES system was $\sim6\times10^{-11}$ mbar, which increased to $\sim2\times10^{-10}$ mbar during the Helium lamp operation. The typical energy and momentum resolution is $\sim$12 meV and $\sim$0.01 {\AA}$^{-1}$, respectively. The RC-FDD function used in Fig. \ref{Fig4}(d) is determined by fitting the temperature-dependent spectra of a polycrystal Au sample, taken under identical experimental conditions as Fig. \ref{Fig4}(c). To compare the spectra from different thicknesses and temperatures in Fig. \ref{Fig4}(d), a renormalization to the background was employed.

\par {To probe the $k_z$ dispersion of electronic states, photon-energy-dependent scans for 3 u.c. and 16 u.c. films were carried out at BL03U in Shanghai Synchrotron Radiation Facility (SSRF). To ensure $in$-$situ$ ARPES measurements, a vaccum suitcase with a base pressure better than $\sim5\times10^{-10}$ mbar were used to transfer the samples grown at Zhejiang University to the BL03U beamline at SSRF. The photon-energy-dependent scans, shown in Fig. S6 in \cite{supplementary}, were performed at $\sim10$ K. }

\subsection{DFT calculations}
\par Electronic structure calculations were performed using DFT as implemented in the Vienna Ab initio Simulation Package (VASP) \cite{PhysRevB.47.558,PhysRevB.59.1758}. The Perdew, Burke and Ernzerhoff (PBE) parametrization of generalized gradient approximation (GGA) to the exchange-correlation functional was employed \cite{PhysRevLett.77.3865}. Calculations that treat Ce $4f$ electrons as core electrons (core-$4f$) or valence electrons (valence-$4f$) were both performed (see {Fig. S3 in \cite{supplementary}}), although only the results from the core-$4f$ calculations were presented in the manuscript for comparison with the experimental data. There are two reasons for this \cite{PhysRevLett.126.216406,Wu_2024}: 1. the many-body Kondo effect modifies the quasiparticle bands only very close to $E_F$ (see Fig. \ref{Fig3}(c)); 2. valence-$4f$ calculations without special treatment of electron correlations often yield incorrect positions of $4f$ bands and overestimate their bandwidth. For all calculations, spin-orbit interaction was considered. An energy cutoff of 400 eV and 9$\times$9$\times$9 (9$\times$9$\times$1 for slab) Gamma-centered K-mesh were employed to converge the total energy to 1 meV/atom. The band structure obtained from the PBE method was fitted to a tight-binding model Hamiltonian with the maximally projected Wannier function method \cite{MOSTOFI2008685}. The resulting Hamiltonian was used to calculate the Fermi surface, as shown in Fig. \ref{Fig2}(b). In slab calculations, a thick CeSi$_{2}$ film with a well-defined surface termination on both sides and a vacuum gap of 18 {\AA} was employed. The calculated bands are shifted upward by 0.2 eV to match with experimental data. The surface weight of the electronic wave function was used to distinguish between the SS and bulk states.

\subsection{$Ex$-$situ$ characterizations}
\par The structure of thin CeSi$_{2}$ films was characterized by $ex$-$situ$ XRD and STEM measurements. The XRD measurements were performed in a Rigaku Ultima IV diffractometer with Cu K$\alpha$ photons (and a small contribution from Cu K$\beta$ photons). The STEM measurements were performed with a state-of-the-art spherical aberration corrected transmission electron microscope, using both the HAADF and iDPC modes.

\par {The $ex$-$situ$ transport measurements were performed in either a Dynacool Quantum-Design Physical Property Measurement System (PPMS) or another PPMS with dilution refrigerator insert, using the standard collinear four-probe method. The absolute value of resistivity $\rho$ was obtained through $\rho =\frac{\pi}{\ln_{}{2} } \frac{V}{I} t\kappa$, where $V$ and $I$ are the measured voltage and current, $t$ is the film thickness and $\kappa$ is a correction factor that depends on the detailed measurement geometry \cite{Miccoli_2015}. For simplicity, we choose $\kappa$ = 1 in our paper, which corresponds to the ideal two-dimensional case. To obtain the magnetic part of the resistivity $\rho_m$($T$) in CeSi$_{2}$ films reliably, we grow epitaxial LaSi$_{2}$ films with the same thicknesses and measure their temperature-dependent resistivity under identical conditions as those of CeSi$_{2}$ films. This ensures identical correction factor $\kappa$ for LaSi$_{2}$ and CeSi$_{2}$ films and hence $\rho_m$($T$) can be reliably obtained via $\rho_m$($T$) = $\rho_{CeSi_{2}}$($T$) - $\rho_{LaSi_{2}}$($T$) (see {Fig. S8 and S11 in \cite{supplementary}}). To minimize possible sample oxidation for $ex$-$situ$ measurements, all the transport data presented in the manuscript were measured immediately after the samples were taken out of the vacuum chamber. We also performed careful control experiments to eliminate the possible complication from sample oxidation (Fig. S12 in \cite{supplementary}). }

\section{ACKNOWLEDGMENTS}
\par This work is supported by the National Key R$\&$D Program of China (Grant No. 2022YFA1402200, No. 2023YFA1406303), the National Science Foundation of China (Grant No. 12525408, No. 12174331) and Zhejiang Provincial Natural Science Foundation of China (Grant No. LGG22E020006, No. LRG26A040001). {We thank Runfeng Zhang, Chufan Chen, Yuping Zhao and Prof. Zhizhan Qiu for help on experiments. We thank the BL03U at SSRF (31124.02.SSRF.BL03U) for the assistance on ARPES measurements.}

\section{DATA AVAILABILITY}
\par The data that support the findings of this study are openly available \cite{CeSi2_figshare}.

%\noindent\textbf{Author contributions}

%The project was designed by Y. L. Thin film growth and $in$-$situ$ ARPES measurement was performed by Y. W., W. Z. and T. H., with help from H. Z.. ARPES data analysis was done by Y. W., W. Z. , T. H. and Y. L. DFT calculations were carried out by Y. F. and C. C. TEM measurements were performed by T. S. $Ex$-$situ$ transport measurements were performed by W. Z., xx and H.-Q. Y. The manuscript was prepared by Y. W., W. Z., T. H., and Y. L., with comments from all coauthors. M. S., F. S. and J. K. contribute to the discussions and date interpretation. All authors discussed the results and commented on the manuscript.

%


\begin{thebibliography}{73}%
\makeatletter
\providecommand \@ifxundefined [1]{%
 \@ifx{#1\undefined}
}%
\providecommand \@ifnum [1]{%
 \ifnum #1\expandafter \@firstoftwo
 \else \expandafter \@secondoftwo
 \fi
}%
\providecommand \@ifx [1]{%
 \ifx #1\expandafter \@firstoftwo
 \else \expandafter \@secondoftwo
 \fi
}%
\providecommand \natexlab [1]{#1}%
\providecommand \enquote  [1]{``#1''}%
\providecommand \bibnamefont  [1]{#1}%
\providecommand \bibfnamefont [1]{#1}%
\providecommand \citenamefont [1]{#1}%
\providecommand \href@noop [0]{\@secondoftwo}%
\providecommand \href [0]{\begingroup \@sanitize@url \@href}%
\providecommand \@href[1]{\@@startlink{#1}\@@href}%
\providecommand \@@href[1]{\endgroup#1\@@endlink}%
\providecommand \@sanitize@url [0]{\catcode `\\12\catcode `\$12\catcode `\&12\catcode `\#12\catcode `\^12\catcode `\_12\catcode `\%12\relax}%
\providecommand \@@startlink[1]{}%
\providecommand \@@endlink[0]{}%
\providecommand \url  [0]{\begingroup\@sanitize@url \@url }%
\providecommand \@url [1]{\endgroup\@href {#1}{\urlprefix }}%
\providecommand \urlprefix  [0]{URL }%
\providecommand \Eprint [0]{\href }%
\providecommand \doibase [0]{https://doi.org/}%
\providecommand \selectlanguage [0]{\@gobble}%
\providecommand \bibinfo  [0]{\@secondoftwo}%
\providecommand \bibfield  [0]{\@secondoftwo}%
\providecommand \translation [1]{[#1]}%
\providecommand \BibitemOpen [0]{}%
\providecommand \bibitemStop [0]{}%
\providecommand \bibitemNoStop [0]{.\EOS\space}%
\providecommand \EOS [0]{\spacefactor3000\relax}%
\providecommand \BibitemShut  [1]{\csname bibitem#1\endcsname}%
\let\auto@bib@innerbib\@empty
%</preamble>
\bibitem [{\citenamefont {Chiang}(2000)}]{ChiangPhotoemission}%
  \BibitemOpen
  \bibfield  {author} {\bibinfo {author} {\bibfnamefont {T.-C.}\ \bibnamefont {Chiang}},\ }\bibfield  {title} {\bibinfo {title} {\textit{Photoemission studies of quantum well states in thin films}},\ }\href {https://doi.org/https://doi.org/10.1016/S0167-5729(00)00006-6} {\bibfield  {journal} {\bibinfo  {journal} {Surface Science Reports}\ }\textbf {\bibinfo {volume} {39}},\ \bibinfo {pages} {181} (\bibinfo {year} {2000})}\BibitemShut {NoStop}%
\bibitem [{\citenamefont {Yoshimatsu}\ \emph {et~al.}(2011)\citenamefont {Yoshimatsu}, \citenamefont {Horiba}, \citenamefont {Kumigashira}, \citenamefont {Yoshida}, \citenamefont {Fujimori},\ and\ \citenamefont {Oshima}}]{Yoshimatsu2011Science}%
  \BibitemOpen
  \bibfield  {author} {\bibinfo {author} {\bibfnamefont {K.}~\bibnamefont {Yoshimatsu}}, \bibinfo {author} {\bibfnamefont {K.}~\bibnamefont {Horiba}}, \bibinfo {author} {\bibfnamefont {H.}~\bibnamefont {Kumigashira}}, \bibinfo {author} {\bibfnamefont {T.}~\bibnamefont {Yoshida}}, \bibinfo {author} {\bibfnamefont {A.}~\bibnamefont {Fujimori}},\ and\ \bibinfo {author} {\bibfnamefont {M.}~\bibnamefont {Oshima}},\ }\bibfield  {title} {\bibinfo {title} {\textit{Metallic Quantum Well States in Artificial Structures of Strongly Correlated Oxide}},\ }\href {https://doi.org/10.1126/science.1205771} {\bibfield  {journal} {\bibinfo  {journal} {Science}\ }\textbf {\bibinfo {volume} {333}},\ \bibinfo {pages} {319} (\bibinfo {year} {2011})}\BibitemShut {NoStop}%
\bibitem [{\citenamefont {Monkman}\ \emph {et~al.}(2012)\citenamefont {Monkman}, \citenamefont {Adamo}, \citenamefont {Mundy}, \citenamefont {Shai}, \citenamefont {Harter}, \citenamefont {Shen}, \citenamefont {Burganov}, \citenamefont {Muller}, \citenamefont {Schlom},\ and\ \citenamefont {Shen}}]{Monkman2012NM}%
  \BibitemOpen
  \bibfield  {author} {\bibinfo {author} {\bibfnamefont {E.~J.}\ \bibnamefont {Monkman}}, \bibinfo {author} {\bibfnamefont {C.}~\bibnamefont {Adamo}}, \bibinfo {author} {\bibfnamefont {J.~A.}\ \bibnamefont {Mundy}}, \bibinfo {author} {\bibfnamefont {D.~E.}\ \bibnamefont {Shai}}, \bibinfo {author} {\bibfnamefont {J.~W.}\ \bibnamefont {Harter}}, \bibinfo {author} {\bibfnamefont {D.}~\bibnamefont {Shen}}, \bibinfo {author} {\bibfnamefont {B.}~\bibnamefont {Burganov}}, \bibinfo {author} {\bibfnamefont {D.~A.}\ \bibnamefont {Muller}}, \bibinfo {author} {\bibfnamefont {D.~G.}\ \bibnamefont {Schlom}},\ and\ \bibinfo {author} {\bibfnamefont {K.~M.}\ \bibnamefont {Shen}},\ }\bibfield  {title} {\bibinfo {title} {\textit{Quantum many-body interactions in digital oxide superlattices}},\ }\href {https://doi.org/10.1038/nmat3405} {\bibfield  {journal} {\bibinfo  {journal} {Nature Materials}\ }\textbf {\bibinfo {volume} {11}},\ \bibinfo {pages} {855} (\bibinfo {year} {2012})}\BibitemShut {NoStop}%
\bibitem [{\citenamefont {King}\ \emph {et~al.}(2014)\citenamefont {King}, \citenamefont {Wei}, \citenamefont {Nie}, \citenamefont {Uchida}, \citenamefont {Adamo}, \citenamefont {Zhu}, \citenamefont {He}, \citenamefont {Bo{\v{z}}ovi{\'{c}}}, \citenamefont {Schlom},\ and\ \citenamefont {Shen}}]{King2014}%
  \BibitemOpen
  \bibfield  {author} {\bibinfo {author} {\bibfnamefont {P.~D.~C.}\ \bibnamefont {King}}, \bibinfo {author} {\bibfnamefont {H.~I.}\ \bibnamefont {Wei}}, \bibinfo {author} {\bibfnamefont {Y.~F.}\ \bibnamefont {Nie}}, \bibinfo {author} {\bibfnamefont {M.}~\bibnamefont {Uchida}}, \bibinfo {author} {\bibfnamefont {C.}~\bibnamefont {Adamo}}, \bibinfo {author} {\bibfnamefont {S.}~\bibnamefont {Zhu}}, \bibinfo {author} {\bibfnamefont {X.}~\bibnamefont {He}}, \bibinfo {author} {\bibfnamefont {I.}~\bibnamefont {Bo{\v{z}}ovi{\'{c}}}}, \bibinfo {author} {\bibfnamefont {D.~G.}\ \bibnamefont {Schlom}},\ and\ \bibinfo {author} {\bibfnamefont {K.~M.}\ \bibnamefont {Shen}},\ }\bibfield  {title} {\bibinfo {title} {\textit{Atomic-scale control of competing electronic phases in ultrathin ${\mathrm{LaNiO}_{3}}$}},\ }\href {https://doi.org/10.1038/nnano.2014.59} {\bibfield  {journal} {\bibinfo  {journal} {Nature Nanotechnology}\ }\textbf {\bibinfo {volume} {9}},\ \bibinfo {pages} {443} (\bibinfo {year} {2014})}\BibitemShut {NoStop}%
\bibitem [{\citenamefont {Coleman}(2007)}]{Coleman2007heavy}%
  \BibitemOpen
  \bibfield  {author} {\bibinfo {author} {\bibfnamefont {P.}~\bibnamefont {Coleman}},\ }\href@noop {} {\emph {\bibinfo {title} {\textit{Heavy Fermions: electrons at the edge of magnetism. Handbook of Magnetism and Advanced Magnetic Materials}}}}\ (\bibinfo  {publisher} {Wiley, New York},\ \bibinfo {year} {2007})\BibitemShut {NoStop}%
\bibitem [{\citenamefont {Gegenwart}\ \emph {et~al.}(2008)\citenamefont {Gegenwart}, \citenamefont {Si},\ and\ \citenamefont {Steglich}}]{gegenwart2008quantum}%
  \BibitemOpen
  \bibfield  {author} {\bibinfo {author} {\bibfnamefont {P.}~\bibnamefont {Gegenwart}}, \bibinfo {author} {\bibfnamefont {Q.}~\bibnamefont {Si}},\ and\ \bibinfo {author} {\bibfnamefont {F.}~\bibnamefont {Steglich}},\ }\bibfield  {title} {\bibinfo {title} {\textit{Quantum criticality in heavy-fermion metals}},\ }\href {https://doi.org/10.1038/nphys892} {\bibfield  {journal} {\bibinfo  {journal} {Nature Physics}\ }\textbf {\bibinfo {volume} {4}},\ \bibinfo {pages} {186} (\bibinfo {year} {2008})}\BibitemShut {NoStop}%
\bibitem [{\citenamefont {Sachdev}(2011)}]{Sachdev2011Quantum}%
  \BibitemOpen
  \bibfield  {author} {\bibinfo {author} {\bibfnamefont {S.}~\bibnamefont {Sachdev}},\ }\href@noop {} {\emph {\bibinfo {title} {\textit{Quantum Phase Transitions}}}}\ (\bibinfo  {publisher} {Cambridge Univ. Press},\ \bibinfo {year} {2011})\BibitemShut {NoStop}%
\bibitem [{\citenamefont {Mathur}\ \emph {et~al.}(1998)\citenamefont {Mathur}, \citenamefont {Grosche}, \citenamefont {Julian}, \citenamefont {Walker}, \citenamefont {Freye}, \citenamefont {Haselwimmer},\ and\ \citenamefont {Lonzarich}}]{Mathur1998Magnetically}%
  \BibitemOpen
  \bibfield  {author} {\bibinfo {author} {\bibfnamefont {N.~D.}\ \bibnamefont {Mathur}}, \bibinfo {author} {\bibfnamefont {F.~M.}\ \bibnamefont {Grosche}}, \bibinfo {author} {\bibfnamefont {S.~R.}\ \bibnamefont {Julian}}, \bibinfo {author} {\bibfnamefont {I.~R.}\ \bibnamefont {Walker}}, \bibinfo {author} {\bibfnamefont {D.~M.}\ \bibnamefont {Freye}}, \bibinfo {author} {\bibfnamefont {R.~K.~W.}\ \bibnamefont {Haselwimmer}},\ and\ \bibinfo {author} {\bibfnamefont {G.~G.}\ \bibnamefont {Lonzarich}},\ }\bibfield  {title} {\bibinfo {title} {\textit{Magnetically mediated superconductivity in heavy fermion compounds}},\ }\href {https://doi.org/10.1038/27838} {\bibfield  {journal} {\bibinfo  {journal} {Nature}\ }\textbf {\bibinfo {volume} {394}},\ \bibinfo {pages} {39} (\bibinfo {year} {1998})}\BibitemShut {NoStop}%
\bibitem [{\citenamefont {L.~Sarrao}\ and\ \citenamefont {D.~Thompson}(2007)}]{JPSJ2007review}%
  \BibitemOpen
  \bibfield  {author} {\bibinfo {author} {\bibfnamefont {J.}~\bibnamefont {L.~Sarrao}}\ and\ \bibinfo {author} {\bibfnamefont {J.}~\bibnamefont {D.~Thompson}},\ }\bibfield  {title} {\bibinfo {title} {\textit{Superconductivity in Cerium- and Plutonium-Based `115' Materials}},\ }\href {https://doi.org/10.1143/JPSJ.76.051013} {\bibfield  {journal} {\bibinfo  {journal} {Journal of the Physical Society of Japan}\ }\textbf {\bibinfo {volume} {76}},\ \bibinfo {pages} {051013} (\bibinfo {year} {2007})}\BibitemShut {NoStop}%
\bibitem [{\citenamefont {L\"ohneysen}\ \emph {et~al.}(2007)\citenamefont {L\"ohneysen}, \citenamefont {Rosch}, \citenamefont {Vojta},\ and\ \citenamefont {W\"olfle}}]{Lohneysen2007Fermi}%
  \BibitemOpen
  \bibfield  {author} {\bibinfo {author} {\bibfnamefont {H.~v.}\ \bibnamefont {L\"ohneysen}}, \bibinfo {author} {\bibfnamefont {A.}~\bibnamefont {Rosch}}, \bibinfo {author} {\bibfnamefont {M.}~\bibnamefont {Vojta}},\ and\ \bibinfo {author} {\bibfnamefont {P.}~\bibnamefont {W\"olfle}},\ }\bibfield  {title} {\bibinfo {title} {\textit{Fermi-liquid instabilities at magnetic quantum phase transitions}},\ }\href {https://doi.org/10.1103/RevModPhys.79.1015} {\bibfield  {journal} {\bibinfo  {journal} {Rev. Mod. Phys.}\ }\textbf {\bibinfo {volume} {79}},\ \bibinfo {pages} {1015} (\bibinfo {year} {2007})}\BibitemShut {NoStop}%
\bibitem [{\citenamefont {White}\ \emph {et~al.}(2015)\citenamefont {White}, \citenamefont {Thompson},\ and\ \citenamefont {Maple}}]{WHITE2015review}%
  \BibitemOpen
  \bibfield  {author} {\bibinfo {author} {\bibfnamefont {B.}~\bibnamefont {White}}, \bibinfo {author} {\bibfnamefont {J.}~\bibnamefont {Thompson}},\ and\ \bibinfo {author} {\bibfnamefont {M.}~\bibnamefont {Maple}},\ }\bibfield  {title} {\bibinfo {title} {\textit{Unconventional superconductivity in heavy-fermion compounds}},\ }\href {https://doi.org/https://doi.org/10.1016/j.physc.2015.02.044} {\bibfield  {journal} {\bibinfo  {journal} {Physica C: Superconductivity and its Applications}\ }\textbf {\bibinfo {volume} {514}},\ \bibinfo {pages} {246} (\bibinfo {year} {2015})}\BibitemShut {NoStop}%
\bibitem [{\citenamefont {Kirchner}\ \emph {et~al.}(2020)\citenamefont {Kirchner}, \citenamefont {Paschen}, \citenamefont {Chen}, \citenamefont {Wirth}, \citenamefont {Feng}, \citenamefont {Thompson},\ and\ \citenamefont {Si}}]{kirchner2020colloquium}%
  \BibitemOpen
  \bibfield  {author} {\bibinfo {author} {\bibfnamefont {S.}~\bibnamefont {Kirchner}}, \bibinfo {author} {\bibfnamefont {S.}~\bibnamefont {Paschen}}, \bibinfo {author} {\bibfnamefont {Q.}~\bibnamefont {Chen}}, \bibinfo {author} {\bibfnamefont {S.}~\bibnamefont {Wirth}}, \bibinfo {author} {\bibfnamefont {D.}~\bibnamefont {Feng}}, \bibinfo {author} {\bibfnamefont {J.~D.}\ \bibnamefont {Thompson}},\ and\ \bibinfo {author} {\bibfnamefont {Q.}~\bibnamefont {Si}},\ }\bibfield  {title} {\bibinfo {title} {\textit{Colloquium: Heavy-electron quantum criticality and single-particle spectroscopy}},\ }\href {https://doi.org/10.1103/RevModPhys.92.011002} {\bibfield  {journal} {\bibinfo  {journal} {Rev. Mod. Phys.}\ }\textbf {\bibinfo {volume} {92}},\ \bibinfo {pages} {011002} (\bibinfo {year} {2020})}\BibitemShut {NoStop}%
\bibitem [{\citenamefont {Smidman}\ \emph {et~al.}(2023)\citenamefont {Smidman}, \citenamefont {Stockert}, \citenamefont {Nica}, \citenamefont {Liu}, \citenamefont {Yuan}, \citenamefont {Si},\ and\ \citenamefont {Steglich}}]{Smidman2023RMP}%
  \BibitemOpen
  \bibfield  {author} {\bibinfo {author} {\bibfnamefont {M.}~\bibnamefont {Smidman}}, \bibinfo {author} {\bibfnamefont {O.}~\bibnamefont {Stockert}}, \bibinfo {author} {\bibfnamefont {E.~M.}\ \bibnamefont {Nica}}, \bibinfo {author} {\bibfnamefont {Y.}~\bibnamefont {Liu}}, \bibinfo {author} {\bibfnamefont {H.}~\bibnamefont {Yuan}}, \bibinfo {author} {\bibfnamefont {Q.}~\bibnamefont {Si}},\ and\ \bibinfo {author} {\bibfnamefont {F.}~\bibnamefont {Steglich}},\ }\bibfield  {title} {\bibinfo {title} {\textit{Colloqium: Unconventional fully gapped superconductivity in the heavy-fermion metal ${\mathrm{CeCu}}_{2}{\mathrm{Si}}_{2}$}},\ }\href {https://doi.org/10.1103/RevModPhys.95.031002} {\bibfield  {journal} {\bibinfo  {journal} {Rev. Mod. Phys.}\ }\textbf {\bibinfo {volume} {95}},\ \bibinfo {pages} {031002} (\bibinfo {year} {2023})}\BibitemShut {NoStop}%
\bibitem [{\citenamefont {Coleman}(2010)}]{coleman2010Science}%
  \BibitemOpen
  \bibfield  {author} {\bibinfo {author} {\bibfnamefont {P.}~\bibnamefont {Coleman}},\ }\bibfield  {title} {\bibinfo {title} {\textit{The Lowdown on Heavy Fermions}},\ }\href {https://doi.org/10.1126/science.1186253} {\bibfield  {journal} {\bibinfo  {journal} {Science}\ }\textbf {\bibinfo {volume} {327}},\ \bibinfo {pages} {969} (\bibinfo {year} {2010})}\BibitemShut {NoStop}%
\bibitem [{\citenamefont {Thompson}(2011)}]{Thompson2011NP}%
  \BibitemOpen
  \bibfield  {author} {\bibinfo {author} {\bibfnamefont {J.~D.}\ \bibnamefont {Thompson}},\ }\bibfield  {title} {\bibinfo {title} {\textit{Heavy electron seeks same}},\ }\href {https://doi.org/10.1038/nphys2129} {\bibfield  {journal} {\bibinfo  {journal} {Nature Physics}\ }\textbf {\bibinfo {volume} {7}},\ \bibinfo {pages} {838} (\bibinfo {year} {2011})}\BibitemShut {NoStop}%
\bibitem [{\citenamefont {Posey}\ \emph {et~al.}(2024)\citenamefont {Posey}, \citenamefont {Turkel}, \citenamefont {Rezaee}, \citenamefont {Devarakonda}, \citenamefont {Kundu}, \citenamefont {Ong}, \citenamefont {Thinel}, \citenamefont {Chica}, \citenamefont {Vitalone}, \citenamefont {Jing}, \citenamefont {Xu}, \citenamefont {Needell}, \citenamefont {Meirzadeh}, \citenamefont {Feuer}, \citenamefont {Jindal}, \citenamefont {Cui}, \citenamefont {Valla}, \citenamefont {Thunstr{\"o}m}, \citenamefont {Yilmaz}, \citenamefont {Vescovo}, \citenamefont {Graf}, \citenamefont {Zhu}, \citenamefont {Scheie}, \citenamefont {May}, \citenamefont {Eriksson}, \citenamefont {Basov}, \citenamefont {Dean}, \citenamefont {Rubio}, \citenamefont {Kim}, \citenamefont {Ziebel}, \citenamefont {Millis}, \citenamefont {Pasupathy},\ and\ \citenamefont {Roy}}]{Posey2024}%
  \BibitemOpen
  \bibfield  {author} {\bibinfo {author} {\bibfnamefont {V.~A.}\ \bibnamefont {Posey}}, \bibinfo {author} {\bibfnamefont {S.}~\bibnamefont {Turkel}}, \bibinfo {author} {\bibfnamefont {M.}~\bibnamefont {Rezaee}}, \bibinfo {author} {\bibfnamefont {A.}~\bibnamefont {Devarakonda}}, \bibinfo {author} {\bibfnamefont {A.~K.}\ \bibnamefont {Kundu}}, \bibinfo {author} {\bibfnamefont {C.~S.}\ \bibnamefont {Ong}}, \bibinfo {author} {\bibfnamefont {M.}~\bibnamefont {Thinel}}, \bibinfo {author} {\bibfnamefont {D.~G.}\ \bibnamefont {Chica}}, \bibinfo {author} {\bibfnamefont {R.~A.}\ \bibnamefont {Vitalone}}, \bibinfo {author} {\bibfnamefont {R.}~\bibnamefont {Jing}}, \bibinfo {author} {\bibfnamefont {S.}~\bibnamefont {Xu}}, \bibinfo {author} {\bibfnamefont {D.~R.}\ \bibnamefont {Needell}}, \bibinfo {author} {\bibfnamefont {E.}~\bibnamefont {Meirzadeh}}, \bibinfo {author} {\bibfnamefont {M.~L.}\ \bibnamefont {Feuer}}, \bibinfo {author} {\bibfnamefont {A.}~\bibnamefont {Jindal}}, \bibinfo {author} {\bibfnamefont {X.}~\bibnamefont {Cui}}, \bibinfo {author} {\bibfnamefont {T.}~\bibnamefont {Valla}}, \bibinfo {author} {\bibfnamefont {P.}~\bibnamefont {Thunstr{\"o}m}}, \bibinfo {author} {\bibfnamefont {T.}~\bibnamefont {Yilmaz}}, \bibinfo {author} {\bibfnamefont {E.}~\bibnamefont {Vescovo}}, \bibinfo {author} {\bibfnamefont {D.}~\bibnamefont {Graf}}, \bibinfo {author} {\bibfnamefont {X.}~\bibnamefont {Zhu}}, \bibinfo {author} {\bibfnamefont {A.}~\bibnamefont {Scheie}}, \bibinfo {author} {\bibfnamefont {A.~F.}\ \bibnamefont {May}}, \bibinfo {author} {\bibfnamefont {O.}~\bibnamefont {Eriksson}}, \bibinfo {author} {\bibfnamefont {D.~N.}\ \bibnamefont {Basov}}, \bibinfo {author} {\bibfnamefont {C.~R.}\ \bibnamefont {Dean}}, \bibinfo {author} {\bibfnamefont {A.}~\bibnamefont {Rubio}}, \bibinfo {author} {\bibfnamefont {P.}~\bibnamefont {Kim}}, \bibinfo {author} {\bibfnamefont {M.~E.}\ \bibnamefont {Ziebel}}, \bibinfo {author} {\bibfnamefont {A.~J.}\ \bibnamefont {Millis}}, \bibinfo {author} {\bibfnamefont {A.~N.}\ \bibnamefont {Pasupathy}},\ and\ \bibinfo {author} {\bibfnamefont {X.}~\bibnamefont {Roy}},\ }\bibfield  {title} {\bibinfo {title} {\textit{Two-dimensional heavy fermions in the van der Waals metal ${\mathrm{CeSiI}}$}},\ }\href {https://doi.org/10.1038/s41586-023-06868-x} {\bibfield  {journal} {\bibinfo  {journal} {Nature}\ }\textbf {\bibinfo {volume} {625}},\ \bibinfo {pages} {483} (\bibinfo {year} {2024})}\BibitemShut {NoStop}%
\bibitem [{\citenamefont {Shishido}\ \emph {et~al.}(2010)\citenamefont {Shishido}, \citenamefont {Shibauchi}, \citenamefont {Yasu}, \citenamefont {Kato}, \citenamefont {Kontani}, \citenamefont {Terashima},\ and\ \citenamefont {Matsuda}}]{Shishido2010Science}%
  \BibitemOpen
  \bibfield  {author} {\bibinfo {author} {\bibfnamefont {H.}~\bibnamefont {Shishido}}, \bibinfo {author} {\bibfnamefont {T.}~\bibnamefont {Shibauchi}}, \bibinfo {author} {\bibfnamefont {K.}~\bibnamefont {Yasu}}, \bibinfo {author} {\bibfnamefont {T.}~\bibnamefont {Kato}}, \bibinfo {author} {\bibfnamefont {H.}~\bibnamefont {Kontani}}, \bibinfo {author} {\bibfnamefont {T.}~\bibnamefont {Terashima}},\ and\ \bibinfo {author} {\bibfnamefont {Y.}~\bibnamefont {Matsuda}},\ }\bibfield  {title} {\bibinfo {title} {\textit{Tuning the Dimensionality of the heavy fermion compound ${\mathrm{CeIn}}_{\mathrm{3}}$ }},\ }\href {https://doi.org/10.1126/science.1183376} {\bibfield  {journal} {\bibinfo  {journal} {Science}\ }\textbf {\bibinfo {volume} {327}},\ \bibinfo {pages} {980} (\bibinfo {year} {2010})}\BibitemShut {NoStop}%
\bibitem [{\citenamefont {Mizukami}\ \emph {et~al.}(2011)\citenamefont {Mizukami}, \citenamefont {Shishido}, \citenamefont {Shibauchi}, \citenamefont {Shimozawa}, \citenamefont {Yasumoto}, \citenamefont {Watanabe}, \citenamefont {Yamashita}, \citenamefont {Ikeda}, \citenamefont {Terashima}, \citenamefont {Kontani},\ and\ \citenamefont {Matsuda}}]{Mizukami2011}%
  \BibitemOpen
  \bibfield  {author} {\bibinfo {author} {\bibfnamefont {Y.}~\bibnamefont {Mizukami}}, \bibinfo {author} {\bibfnamefont {H.}~\bibnamefont {Shishido}}, \bibinfo {author} {\bibfnamefont {T.}~\bibnamefont {Shibauchi}}, \bibinfo {author} {\bibfnamefont {M.}~\bibnamefont {Shimozawa}}, \bibinfo {author} {\bibfnamefont {S.}~\bibnamefont {Yasumoto}}, \bibinfo {author} {\bibfnamefont {D.}~\bibnamefont {Watanabe}}, \bibinfo {author} {\bibfnamefont {M.}~\bibnamefont {Yamashita}}, \bibinfo {author} {\bibfnamefont {H.}~\bibnamefont {Ikeda}}, \bibinfo {author} {\bibfnamefont {T.}~\bibnamefont {Terashima}}, \bibinfo {author} {\bibfnamefont {H.}~\bibnamefont {Kontani}},\ and\ \bibinfo {author} {\bibfnamefont {Y.}~\bibnamefont {Matsuda}},\ }\bibfield  {title} {\bibinfo {title} {\textit{Extremely strong-coupling superconductivity in artificial two-dimensional Kondo lattices}},\ }\href {https://doi.org/10.1038/nphys2112} {\bibfield  {journal} {\bibinfo  {journal} {Nature Physics}\ }\textbf {\bibinfo {volume} {7}},\ \bibinfo {pages} {849} (\bibinfo {year} {2011})}\BibitemShut {NoStop}%
\bibitem [{\citenamefont {Naritsuka}\ \emph {et~al.}(2018)\citenamefont {Naritsuka}, \citenamefont {Rosa}, \citenamefont {Luo}, \citenamefont {Kasahara}, \citenamefont {Tokiwa}, \citenamefont {Ishii}, \citenamefont {Miyake}, \citenamefont {Terashima}, \citenamefont {Shibauchi}, \citenamefont {Ronning}, \citenamefont {Thompson},\ and\ \citenamefont {Matsuda}}]{PhysRevLett.120.187002}%
  \BibitemOpen
  \bibfield  {author} {\bibinfo {author} {\bibfnamefont {M.}~\bibnamefont {Naritsuka}}, \bibinfo {author} {\bibfnamefont {P.~F.~S.}\ \bibnamefont {Rosa}}, \bibinfo {author} {\bibfnamefont {Y.}~\bibnamefont {Luo}}, \bibinfo {author} {\bibfnamefont {Y.}~\bibnamefont {Kasahara}}, \bibinfo {author} {\bibfnamefont {Y.}~\bibnamefont {Tokiwa}}, \bibinfo {author} {\bibfnamefont {T.}~\bibnamefont {Ishii}}, \bibinfo {author} {\bibfnamefont {S.}~\bibnamefont {Miyake}}, \bibinfo {author} {\bibfnamefont {T.}~\bibnamefont {Terashima}}, \bibinfo {author} {\bibfnamefont {T.}~\bibnamefont {Shibauchi}}, \bibinfo {author} {\bibfnamefont {F.}~\bibnamefont {Ronning}}, \bibinfo {author} {\bibfnamefont {J.~D.}\ \bibnamefont {Thompson}},\ and\ \bibinfo {author} {\bibfnamefont {Y.}~\bibnamefont {Matsuda}},\ }\bibfield  {title} {\bibinfo {title} {\textit{Tuning the Pairing Interaction in a $d$-Wave Superconductor by Paramagnons Injected through Interfaces}},\ }\href {https://doi.org/10.1103/PhysRevLett.120.187002} {\bibfield  {journal} {\bibinfo  {journal} {Phys. Rev. Lett.}\ }\textbf {\bibinfo {volume} {120}},\ \bibinfo {pages} {187002} (\bibinfo {year} {2018})}\BibitemShut {NoStop}%
\bibitem [{\citenamefont {Nakamura}\ \emph {et~al.}(2023)\citenamefont {Nakamura}, \citenamefont {Sugihara}, \citenamefont {Chen}, \citenamefont {Yukawa}, \citenamefont {Ohtsubo}, \citenamefont {Tanaka}, \citenamefont {Kitamura}, \citenamefont {Kumigashira},\ and\ \citenamefont {Kimura}}]{Nakamura2023NC}%
  \BibitemOpen
  \bibfield  {author} {\bibinfo {author} {\bibfnamefont {T.}~\bibnamefont {Nakamura}}, \bibinfo {author} {\bibfnamefont {H.}~\bibnamefont {Sugihara}}, \bibinfo {author} {\bibfnamefont {Y.}~\bibnamefont {Chen}}, \bibinfo {author} {\bibfnamefont {R.}~\bibnamefont {Yukawa}}, \bibinfo {author} {\bibfnamefont {Y.}~\bibnamefont {Ohtsubo}}, \bibinfo {author} {\bibfnamefont {K.}~\bibnamefont {Tanaka}}, \bibinfo {author} {\bibfnamefont {M.}~\bibnamefont {Kitamura}}, \bibinfo {author} {\bibfnamefont {H.}~\bibnamefont {Kumigashira}},\ and\ \bibinfo {author} {\bibfnamefont {S.-i.}\ \bibnamefont {Kimura}},\ }\bibfield  {title} {\bibinfo {title} {\textit{Two-dimensional heavy fermion in a monoatomic-layer Kondo lattice ${\mathrm{YbCu}}_{2}$ }},\ }\href {https://doi.org/10.1038/s41467-023-43662-9} {\bibfield  {journal} {\bibinfo  {journal} {Nature Communications}\ }\textbf {\bibinfo {volume} {14}},\ \bibinfo {pages} {7850} (\bibinfo {year} {2023})}\BibitemShut {NoStop}%
\bibitem [{\citenamefont {Cao}\ \emph {et~al.}(2018)\citenamefont {Cao}, \citenamefont {Fatemi}, \citenamefont {Demir}, \citenamefont {Fang}, \citenamefont {Tomarken}, \citenamefont {Luo}, \citenamefont {Sanchez-Yamagishi}, \citenamefont {Watanabe}, \citenamefont {Taniguchi}, \citenamefont {Kaxiras}, \citenamefont {Ashoori},\ and\ \citenamefont {Jarillo-Herrero}}]{Cao2018second}%
  \BibitemOpen
  \bibfield  {author} {\bibinfo {author} {\bibfnamefont {Y.}~\bibnamefont {Cao}}, \bibinfo {author} {\bibfnamefont {V.}~\bibnamefont {Fatemi}}, \bibinfo {author} {\bibfnamefont {A.}~\bibnamefont {Demir}}, \bibinfo {author} {\bibfnamefont {S.}~\bibnamefont {Fang}}, \bibinfo {author} {\bibfnamefont {S.~L.}\ \bibnamefont {Tomarken}}, \bibinfo {author} {\bibfnamefont {J.~Y.}\ \bibnamefont {Luo}}, \bibinfo {author} {\bibfnamefont {J.~D.}\ \bibnamefont {Sanchez-Yamagishi}}, \bibinfo {author} {\bibfnamefont {K.}~\bibnamefont {Watanabe}}, \bibinfo {author} {\bibfnamefont {T.}~\bibnamefont {Taniguchi}}, \bibinfo {author} {\bibfnamefont {E.}~\bibnamefont {Kaxiras}}, \bibinfo {author} {\bibfnamefont {R.~C.}\ \bibnamefont {Ashoori}},\ and\ \bibinfo {author} {\bibfnamefont {P.}~\bibnamefont {Jarillo-Herrero}},\ }\bibfield  {title} {\bibinfo {title} {\textit{Correlated insulator behaviour at half-filling in magic-angle graphene superlattices}},\ }\href {https://doi.org/10.1038/nature26154} {\bibfield  {journal} {\bibinfo  {journal} {Nature}\ }\textbf {\bibinfo {volume} {556}},\ \bibinfo {pages} {80} (\bibinfo {year} {2018})}\BibitemShut {NoStop}%
\bibitem [{\citenamefont {Shi}\ and\ \citenamefont {Dai}(2022)}]{PhysRevB.106.245129}%
  \BibitemOpen
  \bibfield  {author} {\bibinfo {author} {\bibfnamefont {H.}~\bibnamefont {Shi}}\ and\ \bibinfo {author} {\bibfnamefont {X.}~\bibnamefont {Dai}},\ }\bibfield  {title} {\bibinfo {title} {\textit{Heavy-fermion representation for twisted bilayer graphene systems}},\ }\href {https://doi.org/10.1103/PhysRevB.106.245129} {\bibfield  {journal} {\bibinfo  {journal} {Phys. Rev. B}\ }\textbf {\bibinfo {volume} {106}},\ \bibinfo {pages} {245129} (\bibinfo {year} {2022})}\BibitemShut {NoStop}%
\bibitem [{\citenamefont {Song}\ and\ \citenamefont {Bernevig}(2022)}]{PhysRevLett.129.047601}%
  \BibitemOpen
  \bibfield  {author} {\bibinfo {author} {\bibfnamefont {Z.-D.}\ \bibnamefont {Song}}\ and\ \bibinfo {author} {\bibfnamefont {B.~A.}\ \bibnamefont {Bernevig}},\ }\bibfield  {title} {\bibinfo {title} {\textit{Magic-Angle Twisted Bilayer Graphene as a Topological Heavy Fermion Problem}},\ }\href {https://doi.org/10.1103/PhysRevLett.129.047601} {\bibfield  {journal} {\bibinfo  {journal} {Phys. Rev. Lett.}\ }\textbf {\bibinfo {volume} {129}},\ \bibinfo {pages} {047601} (\bibinfo {year} {2022})}\BibitemShut {NoStop}%
\bibitem [{\citenamefont {Va{\v{n}}o}\ \emph {et~al.}(2021)\citenamefont {Va{\v{n}}o}, \citenamefont {Amini}, \citenamefont {Ganguli}, \citenamefont {Chen}, \citenamefont {Lado}, \citenamefont {Kezilebieke},\ and\ \citenamefont {Liljeroth}}]{Vano2021}%
  \BibitemOpen
  \bibfield  {author} {\bibinfo {author} {\bibfnamefont {V.}~\bibnamefont {Va{\v{n}}o}}, \bibinfo {author} {\bibfnamefont {M.}~\bibnamefont {Amini}}, \bibinfo {author} {\bibfnamefont {S.~C.}\ \bibnamefont {Ganguli}}, \bibinfo {author} {\bibfnamefont {G.}~\bibnamefont {Chen}}, \bibinfo {author} {\bibfnamefont {J.~L.}\ \bibnamefont {Lado}}, \bibinfo {author} {\bibfnamefont {S.}~\bibnamefont {Kezilebieke}},\ and\ \bibinfo {author} {\bibfnamefont {P.}~\bibnamefont {Liljeroth}},\ }\bibfield  {title} {\bibinfo {title} {\textit{Artificial heavy fermions in a van der Waals heterostructure}},\ }\href {https://doi.org/10.1038/s41586-021-04021-0} {\bibfield  {journal} {\bibinfo  {journal} {Nature}\ }\textbf {\bibinfo {volume} {599}},\ \bibinfo {pages} {582} (\bibinfo {year} {2021})}\BibitemShut {NoStop}%
\bibitem [{\citenamefont {Fan}\ \emph {et~al.}(2024)\citenamefont {Fan}, \citenamefont {Jin}, \citenamefont {Huang}, \citenamefont {Duan}, \citenamefont {Yu}, \citenamefont {Liu}, \citenamefont {Xia}, \citenamefont {Liu}, \citenamefont {Zhang}, \citenamefont {Xie}, \citenamefont {Tang}, \citenamefont {Chen}, \citenamefont {Zhang}, \citenamefont {Chen}, \citenamefont {Luo}, \citenamefont {Lu}, \citenamefont {Sun},\ and\ \citenamefont {Fu}}]{Fan2024}%
  \BibitemOpen
  \bibfield  {author} {\bibinfo {author} {\bibfnamefont {K.}~\bibnamefont {Fan}}, \bibinfo {author} {\bibfnamefont {H.}~\bibnamefont {Jin}}, \bibinfo {author} {\bibfnamefont {B.}~\bibnamefont {Huang}}, \bibinfo {author} {\bibfnamefont {G.}~\bibnamefont {Duan}}, \bibinfo {author} {\bibfnamefont {R.}~\bibnamefont {Yu}}, \bibinfo {author} {\bibfnamefont {Z.-Y.}\ \bibnamefont {Liu}}, \bibinfo {author} {\bibfnamefont {H.-N.}\ \bibnamefont {Xia}}, \bibinfo {author} {\bibfnamefont {L.-S.}\ \bibnamefont {Liu}}, \bibinfo {author} {\bibfnamefont {Y.}~\bibnamefont {Zhang}}, \bibinfo {author} {\bibfnamefont {T.}~\bibnamefont {Xie}}, \bibinfo {author} {\bibfnamefont {Q.-Y.}\ \bibnamefont {Tang}}, \bibinfo {author} {\bibfnamefont {G.}~\bibnamefont {Chen}}, \bibinfo {author} {\bibfnamefont {W.-H.}\ \bibnamefont {Zhang}}, \bibinfo {author} {\bibfnamefont {F.~C.}\ \bibnamefont {Chen}}, \bibinfo {author} {\bibfnamefont {X.}~\bibnamefont {Luo}}, \bibinfo {author} {\bibfnamefont {W.~J.}\ \bibnamefont {Lu}}, \bibinfo {author} {\bibfnamefont {Y.~P.}\ \bibnamefont {Sun}},\ and\ \bibinfo {author} {\bibfnamefont {Y.-S.}\ \bibnamefont {Fu}},\ }\bibfield  {title} {\bibinfo {title} {\textit{Artificial superconducting Kondo lattice in a van der Waals heterostructure}},\ }\href {https://doi.org/10.1038/s41467-024-53166-9} {\bibfield  {journal} {\bibinfo  {journal} {Nature Communications}\ }\textbf {\bibinfo {volume} {15}},\ \bibinfo {pages} {8797} (\bibinfo {year} {2024})}\BibitemShut {NoStop}%
\bibitem [{\citenamefont {Herrera}\ \emph {et~al.}(2023)\citenamefont {Herrera}, \citenamefont {Guillam{\'o}n}, \citenamefont {Barrena}, \citenamefont {Herrera}, \citenamefont {Galvis}, \citenamefont {Yeyati}, \citenamefont {Rusz}, \citenamefont {Oppeneer}, \citenamefont {Knebel}, \citenamefont {Brison}, \citenamefont {Flouquet}, \citenamefont {Aoki},\ and\ \citenamefont {Suderow}}]{Herrera2023}%
  \BibitemOpen
  \bibfield  {author} {\bibinfo {author} {\bibfnamefont {E.}~\bibnamefont {Herrera}}, \bibinfo {author} {\bibfnamefont {I.}~\bibnamefont {Guillam{\'o}n}}, \bibinfo {author} {\bibfnamefont {V.}~\bibnamefont {Barrena}}, \bibinfo {author} {\bibfnamefont {W.~J.}\ \bibnamefont {Herrera}}, \bibinfo {author} {\bibfnamefont {J.~A.}\ \bibnamefont {Galvis}}, \bibinfo {author} {\bibfnamefont {A.~L.}\ \bibnamefont {Yeyati}}, \bibinfo {author} {\bibfnamefont {J.}~\bibnamefont {Rusz}}, \bibinfo {author} {\bibfnamefont {P.~M.}\ \bibnamefont {Oppeneer}}, \bibinfo {author} {\bibfnamefont {G.}~\bibnamefont {Knebel}}, \bibinfo {author} {\bibfnamefont {J.~P.}\ \bibnamefont {Brison}}, \bibinfo {author} {\bibfnamefont {J.}~\bibnamefont {Flouquet}}, \bibinfo {author} {\bibfnamefont {D.}~\bibnamefont {Aoki}},\ and\ \bibinfo {author} {\bibfnamefont {H.}~\bibnamefont {Suderow}},\ }\bibfield  {title} {\bibinfo {title} {Quantum-well states at the surface of a heavy-fermion superconductor},\ }\href {https://doi.org/10.1038/s41586-023-05830-1} {\bibfield  {journal} {\bibinfo  {journal} {Nature}\ }\textbf {\bibinfo {volume} {616}},\ \bibinfo {pages} {465} (\bibinfo {year} {2023})}\BibitemShut {NoStop}%
\bibitem [{\citenamefont {Chatterjee}\ \emph {et~al.}(2017)\citenamefont {Chatterjee}, \citenamefont {Ruf}, \citenamefont {Wei}, \citenamefont {Finkelstein}, \citenamefont {Schlom},\ and\ \citenamefont {Shen}}]{Chatterjee2017}%
  \BibitemOpen
  \bibfield  {author} {\bibinfo {author} {\bibfnamefont {S.}~\bibnamefont {Chatterjee}}, \bibinfo {author} {\bibfnamefont {J.~P.}\ \bibnamefont {Ruf}}, \bibinfo {author} {\bibfnamefont {H.~I.}\ \bibnamefont {Wei}}, \bibinfo {author} {\bibfnamefont {K.~D.}\ \bibnamefont {Finkelstein}}, \bibinfo {author} {\bibfnamefont {D.~G.}\ \bibnamefont {Schlom}},\ and\ \bibinfo {author} {\bibfnamefont {K.~M.}\ \bibnamefont {Shen}},\ }\bibfield  {title} {\bibinfo {title} {\textit{Lifshitz transition from valence fluctuations in YbAl3}},\ }\href {https://doi.org/10.1038/s41467-017-00946-1} {\bibfield  {journal} {\bibinfo  {journal} {Nature Communications}\ }\textbf {\bibinfo {volume} {8}},\ \bibinfo {pages} {852} (\bibinfo {year} {2017})}\BibitemShut {NoStop}%
\bibitem [{\citenamefont {Galera}\ \emph {et~al.}(1989)\citenamefont {Galera}, \citenamefont {Murani},\ and\ \citenamefont {Pierre}}]{GALERA1989801}%
  \BibitemOpen
  \bibfield  {author} {\bibinfo {author} {\bibfnamefont {R.}~\bibnamefont {Galera}}, \bibinfo {author} {\bibfnamefont {A.}~\bibnamefont {Murani}},\ and\ \bibinfo {author} {\bibfnamefont {J.}~\bibnamefont {Pierre}},\ }\bibfield  {title} {\bibinfo {title} {\textit{Magnetic spectral response in ${\mathrm{CeSi}_{2}}$}},\ }\href {https://doi.org/https://doi.org/10.1016/0921-4526(89)90798-9} {\bibfield  {journal} {\bibinfo  {journal} {Physica B: Condensed Matter}\ }\textbf {\bibinfo {volume} {156-157}},\ \bibinfo {pages} {801} (\bibinfo {year} {1989})}\BibitemShut {NoStop}%
\bibitem [{\citenamefont {Souptel}\ \emph {et~al.}(2004)\citenamefont {Souptel}, \citenamefont {Behr}, \citenamefont {Löser}, \citenamefont {Teresiak}, \citenamefont {Drotziger},\ and\ \citenamefont {Pfleiderer}}]{SOUPTEL2004606}%
  \BibitemOpen
  \bibfield  {author} {\bibinfo {author} {\bibfnamefont {D.}~\bibnamefont {Souptel}}, \bibinfo {author} {\bibfnamefont {G.}~\bibnamefont {Behr}}, \bibinfo {author} {\bibfnamefont {W.}~\bibnamefont {Löser}}, \bibinfo {author} {\bibfnamefont {A.}~\bibnamefont {Teresiak}}, \bibinfo {author} {\bibfnamefont {S.}~\bibnamefont {Drotziger}},\ and\ \bibinfo {author} {\bibfnamefont {C.}~\bibnamefont {Pfleiderer}},\ }\bibfield  {title} {\bibinfo {title} {\textit{$\mathrm{CeSi_{2-\delta}}$ single crystals: growth features and properties}},\ }\href {https://doi.org/https://doi.org/10.1016/j.jcrysgro.2004.04.125} {\bibfield  {journal} {\bibinfo  {journal} {Journal of Crystal Growth}\ }\textbf {\bibinfo {volume} {269}},\ \bibinfo {pages} {606} (\bibinfo {year} {2004})}\BibitemShut {NoStop}%
\bibitem [{\citenamefont {Shin}\ \emph {et~al.}(2024)\citenamefont {Shin}, \citenamefont {Ramires}, \citenamefont {Pomjakushin}, \citenamefont {Plokhikh},\ and\ \citenamefont {Pomjakushina}}]{Shin2023arXiv}%
  \BibitemOpen
  \bibfield  {author} {\bibinfo {author} {\bibfnamefont {S.}~\bibnamefont {Shin}}, \bibinfo {author} {\bibfnamefont {A.}~\bibnamefont {Ramires}}, \bibinfo {author} {\bibfnamefont {V.}~\bibnamefont {Pomjakushin}}, \bibinfo {author} {\bibfnamefont {I.}~\bibnamefont {Plokhikh}},\ and\ \bibinfo {author} {\bibfnamefont {E.}~\bibnamefont {Pomjakushina}},\ }\bibfield  {title} {\bibinfo {title} {\textit{Ferromagnetic quantum critical point protected by nonsymmorphic symmetry in a Kondo metal}},\ }\href {https://doi.org/10.1038/s41467-024-52720-9} {\bibfield  {journal} {\bibinfo  {journal} {Nature Communications}\ }\textbf {\bibinfo {volume} {15}},\ \bibinfo {pages} {8423} (\bibinfo {year} {2024})}\BibitemShut {NoStop}%
\bibitem [{sup()}]{supplementary}%
  \BibitemOpen
  \href@noop {} {}\bibinfo {note} {See online supplementary material at xxx, which includes additional experimental data and DFT calculations.}\BibitemShut {Stop}%
\bibitem [{\citenamefont {Hill}\ \emph {et~al.}(1992)\citenamefont {Hill}, \citenamefont {Willis},\ and\ \citenamefont {Ali}}]{P-Hill1992JPCM}%
  \BibitemOpen
  \bibfield  {author} {\bibinfo {author} {\bibfnamefont {P.}~\bibnamefont {Hill}}, \bibinfo {author} {\bibfnamefont {F.}~\bibnamefont {Willis}},\ and\ \bibinfo {author} {\bibfnamefont {N.}~\bibnamefont {Ali}},\ }\bibfield  {title} {\bibinfo {title} {\textit{Investigation of the magnetic-non-magnetic crossover region in the Kondo lattice system ${\mathrm{CeSi}}_{x}$}},\ }\href {https://doi.org/10.1088/0953-8984/4/21/017} {\bibfield  {journal} {\bibinfo  {journal} {Journal of Physics: Condensed Matter}\ }\textbf {\bibinfo {volume} {4}},\ \bibinfo {pages} {5015} (\bibinfo {year} {1992})}\BibitemShut {NoStop}%
\bibitem [{\citenamefont {Shaheen}\ and\ \citenamefont {Schilling}(1987)}]{PhysRevB.35.6880}%
  \BibitemOpen
  \bibfield  {author} {\bibinfo {author} {\bibfnamefont {S.~A.}\ \bibnamefont {Shaheen}}\ and\ \bibinfo {author} {\bibfnamefont {J.~S.}\ \bibnamefont {Schilling}},\ }\bibfield  {title} {\bibinfo {title} {\textit{Ferromagnetism of ${\mathrm{CeSi}}_{\mathrm{x}}$ at ambient and high pressure}},\ }\href {https://doi.org/10.1103/PhysRevB.35.6880} {\bibfield  {journal} {\bibinfo  {journal} {Phys. Rev. B}\ }\textbf {\bibinfo {volume} {35}},\ \bibinfo {pages} {6880} (\bibinfo {year} {1987})}\BibitemShut {NoStop}%
\bibitem [{\citenamefont {Hellman}\ and\ \citenamefont {Tung}(1988)}]{PhysRevB.37.10786}%
  \BibitemOpen
  \bibfield  {author} {\bibinfo {author} {\bibfnamefont {F.}~\bibnamefont {Hellman}}\ and\ \bibinfo {author} {\bibfnamefont {R.~T.}\ \bibnamefont {Tung}},\ }\bibfield  {title} {\bibinfo {title} {\textit{Surface structure of thin epitaxial ${\mathrm{CoSi}}_{2}$ grown on ${\mathrm{Si(111)}}$}},\ }\href {https://doi.org/10.1103/PhysRevB.37.10786} {\bibfield  {journal} {\bibinfo  {journal} {Phys. Rev. B}\ }\textbf {\bibinfo {volume} {37}},\ \bibinfo {pages} {10786} (\bibinfo {year} {1988})}\BibitemShut {NoStop}%
\bibitem [{\citenamefont {Spence}\ \emph {et~al.}(2000)\citenamefont {Spence}, \citenamefont {Tear}, \citenamefont {Noakes},\ and\ \citenamefont {Bailey}}]{PhysRevB.61.5707}%
  \BibitemOpen
  \bibfield  {author} {\bibinfo {author} {\bibfnamefont {D.~J.}\ \bibnamefont {Spence}}, \bibinfo {author} {\bibfnamefont {S.~P.}\ \bibnamefont {Tear}}, \bibinfo {author} {\bibfnamefont {T.~C.~Q.}\ \bibnamefont {Noakes}},\ and\ \bibinfo {author} {\bibfnamefont {P.}~\bibnamefont {Bailey}},\ }\bibfield  {title} {\bibinfo {title} {\textit{Medium-energy ion scattering studies of two-dimensional rare-earth silicides}},\ }\href {https://doi.org/10.1103/PhysRevB.61.5707} {\bibfield  {journal} {\bibinfo  {journal} {Phys. Rev. B}\ }\textbf {\bibinfo {volume} {61}},\ \bibinfo {pages} {5707} (\bibinfo {year} {2000})}\BibitemShut {NoStop}%
\bibitem [{\citenamefont {Wood}\ \emph {et~al.}(2005)\citenamefont {Wood}, \citenamefont {Bonet}, \citenamefont {Noakes}, \citenamefont {Bailey},\ and\ \citenamefont {Tear}}]{WOOD2005120}%
  \BibitemOpen
  \bibfield  {author} {\bibinfo {author} {\bibfnamefont {T.}~\bibnamefont {Wood}}, \bibinfo {author} {\bibfnamefont {C.}~\bibnamefont {Bonet}}, \bibinfo {author} {\bibfnamefont {T.}~\bibnamefont {Noakes}}, \bibinfo {author} {\bibfnamefont {P.}~\bibnamefont {Bailey}},\ and\ \bibinfo {author} {\bibfnamefont {S.}~\bibnamefont {Tear}},\ }\bibfield  {title} {\bibinfo {title} {\textit{A medium-energy ion scattering investigation of the structure and surface vibrations of two-dimensional ${\mathrm{YSi}}_{2}$ grown on ${\mathrm{Si(111)}}$}},\ }\href {https://doi.org/https://doi.org/10.1016/j.susc.2005.08.032} {\bibfield  {journal} {\bibinfo  {journal} {Surface Science}\ }\textbf {\bibinfo {volume} {598}},\ \bibinfo {pages} {120} (\bibinfo {year} {2005})}\BibitemShut {NoStop}%
\bibitem [{\citenamefont {Fang}\ \emph {et~al.}(2021)\citenamefont {Fang}, \citenamefont {Wang}, \citenamefont {Li}, \citenamefont {Su}, \citenamefont {Le}, \citenamefont {Wu}, \citenamefont {Yang}, \citenamefont {Zhang}, \citenamefont {Xiao}, \citenamefont {Sun}, \citenamefont {Hong}, \citenamefont {Xie}, \citenamefont {Wang}, \citenamefont {Cao}, \citenamefont {Lu}, \citenamefont {Yuan},\ and\ \citenamefont {Liu}}]{Fang_2021}%
  \BibitemOpen
  \bibfield  {author} {\bibinfo {author} {\bibfnamefont {Y.}~\bibnamefont {Fang}}, \bibinfo {author} {\bibfnamefont {D.}~\bibnamefont {Wang}}, \bibinfo {author} {\bibfnamefont {P.}~\bibnamefont {Li}}, \bibinfo {author} {\bibfnamefont {H.}~\bibnamefont {Su}}, \bibinfo {author} {\bibfnamefont {T.}~\bibnamefont {Le}}, \bibinfo {author} {\bibfnamefont {Y.}~\bibnamefont {Wu}}, \bibinfo {author} {\bibfnamefont {G.-W.}\ \bibnamefont {Yang}}, \bibinfo {author} {\bibfnamefont {H.-L.}\ \bibnamefont {Zhang}}, \bibinfo {author} {\bibfnamefont {Z.-G.}\ \bibnamefont {Xiao}}, \bibinfo {author} {\bibfnamefont {Y.-Q.}\ \bibnamefont {Sun}}, \bibinfo {author} {\bibfnamefont {S.-Y.}\ \bibnamefont {Hong}}, \bibinfo {author} {\bibfnamefont {Y.-W.}\ \bibnamefont {Xie}}, \bibinfo {author} {\bibfnamefont {H.-H.}\ \bibnamefont {Wang}}, \bibinfo {author} {\bibfnamefont {C.}~\bibnamefont {Cao}}, \bibinfo {author} {\bibfnamefont {X.}~\bibnamefont {Lu}}, \bibinfo {author} {\bibfnamefont {H.-Q.}\ \bibnamefont {Yuan}},\ and\ \bibinfo {author} {\bibfnamefont {Y.}~\bibnamefont {Liu}},\ }\bibfield  {title} {\bibinfo {title} {\textit{Growth, electronic structure and superconductivity of ultrathin epitaxial ${\mathrm{CoSi}_{2}}$ films}},\ }\href {https://doi.org/10.1088/1361-648X/abdff6} {\bibfield  {journal} {\bibinfo  {journal} {Journal of Physics: Condensed Matter}\ }\textbf {\bibinfo {volume} {33}},\ \bibinfo {pages} {155501} (\bibinfo {year} {2021})}\BibitemShut {NoStop}%
\bibitem [{\citenamefont {Patthey}\ \emph {et~al.}(1987)\citenamefont {Patthey}, \citenamefont {Schneider}, \citenamefont {Baer},\ and\ \citenamefont {Delley}}]{PhysRevLett.58.2810}%
  \BibitemOpen
  \bibfield  {author} {\bibinfo {author} {\bibfnamefont {F.}~\bibnamefont {Patthey}}, \bibinfo {author} {\bibfnamefont {W.~D.}\ \bibnamefont {Schneider}}, \bibinfo {author} {\bibfnamefont {Y.}~\bibnamefont {Baer}},\ and\ \bibinfo {author} {\bibfnamefont {B.}~\bibnamefont {Delley}},\ }\bibfield  {title} {\bibinfo {title} {\textit{High-temperature collapse of the Kondo resonance in ${{\mathrm{CeSi}}_{2}}$ observed by photoemission}},\ }\href {https://doi.org/10.1103/PhysRevLett.58.2810} {\bibfield  {journal} {\bibinfo  {journal} {Phys. Rev. Lett.}\ }\textbf {\bibinfo {volume} {58}},\ \bibinfo {pages} {2810} (\bibinfo {year} {1987})}\BibitemShut {NoStop}%
\bibitem [{\citenamefont {Allen}(2005)}]{jw2005kondo}%
  \BibitemOpen
  \bibfield  {author} {\bibinfo {author} {\bibfnamefont {J.~W.}\ \bibnamefont {Allen}},\ }\bibfield  {title} {\bibinfo {title} {\textit{The Kondo resonance in electron spectroscopy}},\ }\href {https://doi.org/10.1143/JPSJ.74.34} {\bibfield  {journal} {\bibinfo  {journal} {Journal of the Physical Society of Japan}\ }\textbf {\bibinfo {volume} {74}},\ \bibinfo {pages} {34} (\bibinfo {year} {2005})}\BibitemShut {NoStop}%
\bibitem [{\citenamefont {Fujimori}(2016)}]{fujimori2016band}%
  \BibitemOpen
  \bibfield  {author} {\bibinfo {author} {\bibfnamefont {S.-i.}\ \bibnamefont {Fujimori}},\ }\bibfield  {title} {\bibinfo {title} {\textit{Band structures of $4f$ and $5f$ materials studied by angle-resolved photoelectron spectroscopy}},\ }\href {https://doi.org/10.1088/0953-8984/28/15/153002} {\bibfield  {journal} {\bibinfo  {journal} {Journal of Physics: Condensed Matter}\ }\textbf {\bibinfo {volume} {28}},\ \bibinfo {pages} {153002} (\bibinfo {year} {2016})}\BibitemShut {NoStop}%
\bibitem [{\citenamefont {Denlinger}\ \emph {et~al.}(2001)\citenamefont {Denlinger}, \citenamefont {Gweon}, \citenamefont {Allen}, \citenamefont {Olson}, \citenamefont {Maple}, \citenamefont {Sarrao}, \citenamefont {Armstrong}, \citenamefont {Fisk},\ and\ \citenamefont {Yamagami}}]{denlinger2001comparative}%
  \BibitemOpen
  \bibfield  {author} {\bibinfo {author} {\bibfnamefont {J.~D.}\ \bibnamefont {Denlinger}}, \bibinfo {author} {\bibfnamefont {G.-H.}\ \bibnamefont {Gweon}}, \bibinfo {author} {\bibfnamefont {J.~W.}\ \bibnamefont {Allen}}, \bibinfo {author} {\bibfnamefont {C.~G.}\ \bibnamefont {Olson}}, \bibinfo {author} {\bibfnamefont {M.~B.}\ \bibnamefont {Maple}}, \bibinfo {author} {\bibfnamefont {J.}~\bibnamefont {Sarrao}}, \bibinfo {author} {\bibfnamefont {P.}~\bibnamefont {Armstrong}}, \bibinfo {author} {\bibfnamefont {Z.}~\bibnamefont {Fisk}},\ and\ \bibinfo {author} {\bibfnamefont {H.}~\bibnamefont {Yamagami}},\ }\bibfield  {title} {\bibinfo {title} {\textit{Comparative study of the electronic structure of ${\mathrm{XRu}}_{2}{\mathrm{Si}}_{2}$: probing the Anderson lattice}},\ }\href {https://doi.org/10.1016/S0368-2048(01)00257-2} {\bibfield  {journal} {\bibinfo  {journal} {Journal of Electron Spectroscopy and Related Phenomena}\ }\textbf {\bibinfo {volume} {117}},\ \bibinfo {pages} {347} (\bibinfo {year} {2001})}\BibitemShut {NoStop}%
\bibitem [{\citenamefont {Kummer}\ \emph {et~al.}(2015)\citenamefont {Kummer}, \citenamefont {Patil}, \citenamefont {Chikina}, \citenamefont {G\"uttler}, \citenamefont {H\"oppner}, \citenamefont {Generalov}, \citenamefont {Danzenb\"acher}, \citenamefont {Seiro}, \citenamefont {Hannaske}, \citenamefont {Krellner}, \citenamefont {Kucherenko}, \citenamefont {Shi}, \citenamefont {Radovic}, \citenamefont {Rienks}, \citenamefont {Zwicknagl}, \citenamefont {Matho}, \citenamefont {Allen}, \citenamefont {Laubschat}, \citenamefont {Geibel},\ and\ \citenamefont {Vyalikh}}]{PhysRevX.5.011028}%
  \BibitemOpen
  \bibfield  {author} {\bibinfo {author} {\bibfnamefont {K.}~\bibnamefont {Kummer}}, \bibinfo {author} {\bibfnamefont {S.}~\bibnamefont {Patil}}, \bibinfo {author} {\bibfnamefont {A.}~\bibnamefont {Chikina}}, \bibinfo {author} {\bibfnamefont {M.}~\bibnamefont {G\"uttler}}, \bibinfo {author} {\bibfnamefont {M.}~\bibnamefont {H\"oppner}}, \bibinfo {author} {\bibfnamefont {A.}~\bibnamefont {Generalov}}, \bibinfo {author} {\bibfnamefont {S.}~\bibnamefont {Danzenb\"acher}}, \bibinfo {author} {\bibfnamefont {S.}~\bibnamefont {Seiro}}, \bibinfo {author} {\bibfnamefont {A.}~\bibnamefont {Hannaske}}, \bibinfo {author} {\bibfnamefont {C.}~\bibnamefont {Krellner}}, \bibinfo {author} {\bibfnamefont {Y.}~\bibnamefont {Kucherenko}}, \bibinfo {author} {\bibfnamefont {M.}~\bibnamefont {Shi}}, \bibinfo {author} {\bibfnamefont {M.}~\bibnamefont {Radovic}}, \bibinfo {author} {\bibfnamefont {E.}~\bibnamefont {Rienks}}, \bibinfo {author} {\bibfnamefont {G.}~\bibnamefont {Zwicknagl}}, \bibinfo {author} {\bibfnamefont {K.}~\bibnamefont {Matho}}, \bibinfo {author} {\bibfnamefont {J.~W.}\ \bibnamefont {Allen}}, \bibinfo {author} {\bibfnamefont {C.}~\bibnamefont {Laubschat}}, \bibinfo {author} {\bibfnamefont {C.}~\bibnamefont {Geibel}},\ and\ \bibinfo {author} {\bibfnamefont {D.~V.}\ \bibnamefont {Vyalikh}},\ }\bibfield  {title} {\bibinfo {title} {\textit{Temperature-Independent Fermi Surface in the Kondo Lattice ${\mathrm{YbRh}}_{2}{\mathrm{Si}}_{2}$}},\ }\href {https://doi.org/10.1103/PhysRevX.5.011028} {\bibfield  {journal} {\bibinfo  {journal} {Phys. Rev. X}\ }\textbf {\bibinfo {volume} {5}},\ \bibinfo {pages} {011028} (\bibinfo {year} {2015})}\BibitemShut {NoStop}%
\bibitem [{\citenamefont {Patil}\ \emph {et~al.}(2016)\citenamefont {Patil}, \citenamefont {Generalov}, \citenamefont {G\"uttler}, \citenamefont {Kushwaha}, \citenamefont {Chikina}, \citenamefont {Kummer}, \citenamefont {R\"odel}, \citenamefont {Santander-Syro}, \citenamefont {Caroca-Canales}, \citenamefont {Geibel}, \citenamefont {Danzenb\"acher}, \citenamefont {Kucherenko}, \citenamefont {Laubschat}, \citenamefont {Allen},\ and\ \citenamefont {Vyalikh}}]{patil2016arpes}%
  \BibitemOpen
  \bibfield  {author} {\bibinfo {author} {\bibfnamefont {S.}~\bibnamefont {Patil}}, \bibinfo {author} {\bibfnamefont {A.}~\bibnamefont {Generalov}}, \bibinfo {author} {\bibfnamefont {M.}~\bibnamefont {G\"uttler}}, \bibinfo {author} {\bibfnamefont {P.}~\bibnamefont {Kushwaha}}, \bibinfo {author} {\bibfnamefont {A.}~\bibnamefont {Chikina}}, \bibinfo {author} {\bibfnamefont {K.}~\bibnamefont {Kummer}}, \bibinfo {author} {\bibfnamefont {T.~C.}\ \bibnamefont {R\"odel}}, \bibinfo {author} {\bibfnamefont {A.~F.}\ \bibnamefont {Santander-Syro}}, \bibinfo {author} {\bibfnamefont {N.}~\bibnamefont {Caroca-Canales}}, \bibinfo {author} {\bibfnamefont {C.}~\bibnamefont {Geibel}}, \bibinfo {author} {\bibfnamefont {S.}~\bibnamefont {Danzenb\"acher}}, \bibinfo {author} {\bibfnamefont {Y.}~\bibnamefont {Kucherenko}}, \bibinfo {author} {\bibfnamefont {C.}~\bibnamefont {Laubschat}}, \bibinfo {author} {\bibfnamefont {J.~W.}\ \bibnamefont {Allen}},\ and\ \bibinfo {author} {\bibfnamefont {D.~V.}\ \bibnamefont {Vyalikh}},\ }\bibfield  {title} {\bibinfo {title} {\textit{ARPES view on surface and bulk hybridization phenomena in the antiferromagnetic Kondo lattice ${{\mathrm{CeRh}}_{2}{\mathrm{Si}}_{2}}$ }},\ }\href {https://doi.org/10.1038/ncomms11029} {\bibfield  {journal} {\bibinfo  {journal} {Nature Communications}\ }\textbf {\bibinfo {volume} {7}},\ \bibinfo {pages} {1} (\bibinfo {year} {2016})}\BibitemShut {NoStop}%
\bibitem [{\citenamefont {Chen}\ \emph {et~al.}(2017)\citenamefont {Chen}, \citenamefont {Xu}, \citenamefont {Niu}, \citenamefont {Jiang}, \citenamefont {Peng}, \citenamefont {Xu}, \citenamefont {Wen}, \citenamefont {Ding}, \citenamefont {Huang}, \citenamefont {Shu}, \citenamefont {Zhang}, \citenamefont {Lee}, \citenamefont {Strocov}, \citenamefont {Shi}, \citenamefont {Bisti}, \citenamefont {Schmitt}, \citenamefont {Huang}, \citenamefont {Dudin}, \citenamefont {Lai}, \citenamefont {Kirchner}, \citenamefont {Yuan},\ and\ \citenamefont {Feng}}]{Chen2017}%
  \BibitemOpen
  \bibfield  {author} {\bibinfo {author} {\bibfnamefont {Q.~Y.}\ \bibnamefont {Chen}}, \bibinfo {author} {\bibfnamefont {D.~F.}\ \bibnamefont {Xu}}, \bibinfo {author} {\bibfnamefont {X.~H.}\ \bibnamefont {Niu}}, \bibinfo {author} {\bibfnamefont {J.}~\bibnamefont {Jiang}}, \bibinfo {author} {\bibfnamefont {R.}~\bibnamefont {Peng}}, \bibinfo {author} {\bibfnamefont {H.~C.}\ \bibnamefont {Xu}}, \bibinfo {author} {\bibfnamefont {C.~H.~P.}\ \bibnamefont {Wen}}, \bibinfo {author} {\bibfnamefont {Z.~F.}\ \bibnamefont {Ding}}, \bibinfo {author} {\bibfnamefont {K.}~\bibnamefont {Huang}}, \bibinfo {author} {\bibfnamefont {L.}~\bibnamefont {Shu}}, \bibinfo {author} {\bibfnamefont {Y.~J.}\ \bibnamefont {Zhang}}, \bibinfo {author} {\bibfnamefont {H.}~\bibnamefont {Lee}}, \bibinfo {author} {\bibfnamefont {V.~N.}\ \bibnamefont {Strocov}}, \bibinfo {author} {\bibfnamefont {M.}~\bibnamefont {Shi}}, \bibinfo {author} {\bibfnamefont {F.}~\bibnamefont {Bisti}}, \bibinfo {author} {\bibfnamefont {T.}~\bibnamefont {Schmitt}}, \bibinfo {author} {\bibfnamefont {Y.~B.}\ \bibnamefont {Huang}}, \bibinfo {author} {\bibfnamefont {P.}~\bibnamefont {Dudin}}, \bibinfo {author} {\bibfnamefont {X.~C.}\ \bibnamefont {Lai}}, \bibinfo {author} {\bibfnamefont {S.}~\bibnamefont {Kirchner}}, \bibinfo {author} {\bibfnamefont {H.~Q.}\ \bibnamefont {Yuan}},\ and\ \bibinfo {author} {\bibfnamefont {D.~L.}\ \bibnamefont {Feng}},\ }\bibfield  {title} {\bibinfo {title} {\textit{Direct observation of how the heavy-fermion state develops in ${\mathrm{CeCoIn}}_{5}$}},\ }\href {https://doi.org/10.1103/PhysRevB.96.045107} {\bibfield  {journal} {\bibinfo  {journal} {Phys. Rev. B}\ }\textbf {\bibinfo {volume} {96}},\ \bibinfo {pages} {045107} (\bibinfo {year} {2017})}\BibitemShut {NoStop}%
\bibitem [{\citenamefont {Chen}\ \emph {et~al.}(2018{\natexlab{a}})\citenamefont {Chen}, \citenamefont {Xu}, \citenamefont {Niu}, \citenamefont {Peng}, \citenamefont {Xu}, \citenamefont {Wen}, \citenamefont {Liu}, \citenamefont {Shu}, \citenamefont {Tan}, \citenamefont {Lai}, \citenamefont {Zhang}, \citenamefont {Lee}, \citenamefont {Strocov}, \citenamefont {Bisti}, \citenamefont {Dudin}, \citenamefont {Zhu}, \citenamefont {Yuan}, \citenamefont {Kirchner},\ and\ \citenamefont {Feng}}]{Chen2018}%
  \BibitemOpen
  \bibfield  {author} {\bibinfo {author} {\bibfnamefont {Q.~Y.}\ \bibnamefont {Chen}}, \bibinfo {author} {\bibfnamefont {D.~F.}\ \bibnamefont {Xu}}, \bibinfo {author} {\bibfnamefont {X.~H.}\ \bibnamefont {Niu}}, \bibinfo {author} {\bibfnamefont {R.}~\bibnamefont {Peng}}, \bibinfo {author} {\bibfnamefont {H.~C.}\ \bibnamefont {Xu}}, \bibinfo {author} {\bibfnamefont {C.~H.~P.}\ \bibnamefont {Wen}}, \bibinfo {author} {\bibfnamefont {X.}~\bibnamefont {Liu}}, \bibinfo {author} {\bibfnamefont {L.}~\bibnamefont {Shu}}, \bibinfo {author} {\bibfnamefont {S.~Y.}\ \bibnamefont {Tan}}, \bibinfo {author} {\bibfnamefont {X.~C.}\ \bibnamefont {Lai}}, \bibinfo {author} {\bibfnamefont {Y.~J.}\ \bibnamefont {Zhang}}, \bibinfo {author} {\bibfnamefont {H.}~\bibnamefont {Lee}}, \bibinfo {author} {\bibfnamefont {V.~N.}\ \bibnamefont {Strocov}}, \bibinfo {author} {\bibfnamefont {F.}~\bibnamefont {Bisti}}, \bibinfo {author} {\bibfnamefont {P.}~\bibnamefont {Dudin}}, \bibinfo {author} {\bibfnamefont {J.-X.}\ \bibnamefont {Zhu}}, \bibinfo {author} {\bibfnamefont {H.~Q.}\ \bibnamefont {Yuan}}, \bibinfo {author} {\bibfnamefont {S.}~\bibnamefont {Kirchner}},\ and\ \bibinfo {author} {\bibfnamefont {D.~L.}\ \bibnamefont {Feng}},\ }\bibfield  {title} {\bibinfo {title} {\textit{Band Dependent Interlayer $f$-Electron Hybridization in ${\mathrm{CeRhIn}}_{5}$}},\ }\href {https://doi.org/10.1103/PhysRevLett.120.066403} {\bibfield  {journal} {\bibinfo  {journal} {Phys. Rev. Lett.}\ }\textbf {\bibinfo {volume} {120}},\ \bibinfo {pages} {066403} (\bibinfo {year} {2018}{\natexlab{a}})}\BibitemShut {NoStop}%
\bibitem [{\citenamefont {Jang}\ \emph {et~al.}(2020)\citenamefont {Jang}, \citenamefont {Denlinger}, \citenamefont {Allen}, \citenamefont {Zapf}, \citenamefont {Maple}, \citenamefont {Kim}, \citenamefont {Jang},\ and\ \citenamefont {Shim}}]{Jang2020}%
  \BibitemOpen
  \bibfield  {author} {\bibinfo {author} {\bibfnamefont {S.~Y.}\ \bibnamefont {Jang}}, \bibinfo {author} {\bibfnamefont {J.~D.}\ \bibnamefont {Denlinger}}, \bibinfo {author} {\bibfnamefont {J.~W.}\ \bibnamefont {Allen}}, \bibinfo {author} {\bibfnamefont {V.~S.}\ \bibnamefont {Zapf}}, \bibinfo {author} {\bibfnamefont {M.~B.}\ \bibnamefont {Maple}}, \bibinfo {author} {\bibfnamefont {J.~N.}\ \bibnamefont {Kim}}, \bibinfo {author} {\bibfnamefont {B.~G.}\ \bibnamefont {Jang}},\ and\ \bibinfo {author} {\bibfnamefont {J.~H.}\ \bibnamefont {Shim}},\ }\bibfield  {title} {\bibinfo {title} {\textit{Evolution of the Kondo lattice electronic structure above the transport coherence temperature}},\ }\href {https://doi.org/10.1073/pnas.2001778117} {\bibfield  {journal} {\bibinfo  {journal} {Proc. Natl. Acad. Sci.}\ }\textbf {\bibinfo {volume} {117}},\ \bibinfo {pages} {23467} (\bibinfo {year} {2020})}\BibitemShut {NoStop}%
\bibitem [{\citenamefont {Chen}\ \emph {et~al.}(2024)\citenamefont {Chen}, \citenamefont {Wang}, \citenamefont {Ishizuka}, \citenamefont {Zhang}, \citenamefont {Nogaki}, \citenamefont {Cheng}, \citenamefont {Yang}, \citenamefont {Chen}, \citenamefont {Zhu}, \citenamefont {Liu}, \citenamefont {Mei}, \citenamefont {Yanase}, \citenamefont {Lv},\ and\ \citenamefont {Huang}}]{PhysRevX.14.021048}%
  \BibitemOpen
  \bibfield  {author} {\bibinfo {author} {\bibfnamefont {X.}~\bibnamefont {Chen}}, \bibinfo {author} {\bibfnamefont {L.}~\bibnamefont {Wang}}, \bibinfo {author} {\bibfnamefont {J.}~\bibnamefont {Ishizuka}}, \bibinfo {author} {\bibfnamefont {R.}~\bibnamefont {Zhang}}, \bibinfo {author} {\bibfnamefont {K.}~\bibnamefont {Nogaki}}, \bibinfo {author} {\bibfnamefont {Y.}~\bibnamefont {Cheng}}, \bibinfo {author} {\bibfnamefont {F.}~\bibnamefont {Yang}}, \bibinfo {author} {\bibfnamefont {Z.}~\bibnamefont {Chen}}, \bibinfo {author} {\bibfnamefont {F.}~\bibnamefont {Zhu}}, \bibinfo {author} {\bibfnamefont {Z.}~\bibnamefont {Liu}}, \bibinfo {author} {\bibfnamefont {J.}~\bibnamefont {Mei}}, \bibinfo {author} {\bibfnamefont {Y.}~\bibnamefont {Yanase}}, \bibinfo {author} {\bibfnamefont {B.}~\bibnamefont {Lv}},\ and\ \bibinfo {author} {\bibfnamefont {Y.}~\bibnamefont {Huang}},\ }\bibfield  {title} {\bibinfo {title} {\textit{Coexistence of near-${E}_{F}$ Flat Band and Van Hove Singularity in a Two-Phase Superconductor}},\ }\href {https://doi.org/10.1103/PhysRevX.14.021048} {\bibfield  {journal} {\bibinfo  {journal} {Phys. Rev. X}\ }\textbf {\bibinfo {volume} {14}},\ \bibinfo {pages} {021048} (\bibinfo {year} {2024})}\BibitemShut {NoStop}%
\bibitem [{\citenamefont {Kroha}\ \emph {et~al.}(2003)\citenamefont {Kroha}, \citenamefont {Kirchner}, \citenamefont {Sellier}, \citenamefont {Wölfle}, \citenamefont {Ehm}, \citenamefont {Reinert}, \citenamefont {Hüfner},\ and\ \citenamefont {Geibel}}]{KROHA200369}%
  \BibitemOpen
  \bibfield  {author} {\bibinfo {author} {\bibfnamefont {J.}~\bibnamefont {Kroha}}, \bibinfo {author} {\bibfnamefont {S.}~\bibnamefont {Kirchner}}, \bibinfo {author} {\bibfnamefont {G.}~\bibnamefont {Sellier}}, \bibinfo {author} {\bibfnamefont {P.}~\bibnamefont {Wölfle}}, \bibinfo {author} {\bibfnamefont {D.}~\bibnamefont {Ehm}}, \bibinfo {author} {\bibfnamefont {F.}~\bibnamefont {Reinert}}, \bibinfo {author} {\bibfnamefont {S.}~\bibnamefont {Hüfner}},\ and\ \bibinfo {author} {\bibfnamefont {C.}~\bibnamefont {Geibel}},\ }\bibfield  {title} {\bibinfo {title} {\textit{Structure and transport in multi-orbital Kondo systems}},\ }\href {https://doi.org/https://doi.org/10.1016/S1386-9477(02)00977-3} {\bibfield  {journal} {\bibinfo  {journal} {Physica E: Low-dimensional Systems and Nanostructures}\ }\textbf {\bibinfo {volume} {18}},\ \bibinfo {pages} {69} (\bibinfo {year} {2003})},\ \bibinfo {note} {23rd International Conference on Low Temperature Physics (LT23)}\BibitemShut {NoStop}%
\bibitem [{\citenamefont {Ehm}\ \emph {et~al.}(2007)\citenamefont {Ehm}, \citenamefont {H\"ufner}, \citenamefont {Reinert}, \citenamefont {Kroha}, \citenamefont {W\"olfle}, \citenamefont {Stockert}, \citenamefont {Geibel},\ and\ \citenamefont {L\"ohneysen}}]{ehm2007high}%
  \BibitemOpen
  \bibfield  {author} {\bibinfo {author} {\bibfnamefont {D.}~\bibnamefont {Ehm}}, \bibinfo {author} {\bibfnamefont {S.}~\bibnamefont {H\"ufner}}, \bibinfo {author} {\bibfnamefont {F.}~\bibnamefont {Reinert}}, \bibinfo {author} {\bibfnamefont {J.}~\bibnamefont {Kroha}}, \bibinfo {author} {\bibfnamefont {P.}~\bibnamefont {W\"olfle}}, \bibinfo {author} {\bibfnamefont {O.}~\bibnamefont {Stockert}}, \bibinfo {author} {\bibfnamefont {C.}~\bibnamefont {Geibel}},\ and\ \bibinfo {author} {\bibfnamefont {H.~v.}\ \bibnamefont {L\"ohneysen}},\ }\bibfield  {title} {\bibinfo {title} {\textit{High-resolution photoemission study on low-${T}_{K}$ Ce systems: Kondo resonance, crystal field structures, and their temperature dependence}},\ }\href {https://doi.org/10.1103/PhysRevB.76.045117} {\bibfield  {journal} {\bibinfo  {journal} {Phys. Rev. B}\ }\textbf {\bibinfo {volume} {76}},\ \bibinfo {pages} {045117} (\bibinfo {year} {2007})}\BibitemShut {NoStop}%
\bibitem [{\citenamefont {Chen}\ \emph {et~al.}(2018{\natexlab{b}})\citenamefont {Chen}, \citenamefont {Wen}, \citenamefont {Yao}, \citenamefont {Huang}, \citenamefont {Ding}, \citenamefont {Shu}, \citenamefont {Niu}, \citenamefont {Zhang}, \citenamefont {Lai}, \citenamefont {Huang}, \citenamefont {Zhang}, \citenamefont {Kirchner},\ and\ \citenamefont {Feng}}]{PhysRevB.97.075149}%
  \BibitemOpen
  \bibfield  {author} {\bibinfo {author} {\bibfnamefont {Q.~Y.}\ \bibnamefont {Chen}}, \bibinfo {author} {\bibfnamefont {C.~H.~P.}\ \bibnamefont {Wen}}, \bibinfo {author} {\bibfnamefont {Q.}~\bibnamefont {Yao}}, \bibinfo {author} {\bibfnamefont {K.}~\bibnamefont {Huang}}, \bibinfo {author} {\bibfnamefont {Z.~F.}\ \bibnamefont {Ding}}, \bibinfo {author} {\bibfnamefont {L.}~\bibnamefont {Shu}}, \bibinfo {author} {\bibfnamefont {X.~H.}\ \bibnamefont {Niu}}, \bibinfo {author} {\bibfnamefont {Y.}~\bibnamefont {Zhang}}, \bibinfo {author} {\bibfnamefont {X.~C.}\ \bibnamefont {Lai}}, \bibinfo {author} {\bibfnamefont {Y.~B.}\ \bibnamefont {Huang}}, \bibinfo {author} {\bibfnamefont {G.~B.}\ \bibnamefont {Zhang}}, \bibinfo {author} {\bibfnamefont {S.}~\bibnamefont {Kirchner}},\ and\ \bibinfo {author} {\bibfnamefont {D.~L.}\ \bibnamefont {Feng}},\ }\bibfield  {title} {\bibinfo {title} {\textit{Tracing crystal-field splittings in the rare-earth-based intermetallic ${\mathrm{CeIrIn}}_{5}$}},\ }\href {https://doi.org/10.1103/PhysRevB.97.075149} {\bibfield  {journal} {\bibinfo  {journal} {Phys. Rev. B}\ }\textbf {\bibinfo {volume} {97}},\ \bibinfo {pages} {075149} (\bibinfo {year} {2018}{\natexlab{b}})}\BibitemShut {NoStop}%
\bibitem [{\citenamefont {Pal}\ \emph {et~al.}(2019)\citenamefont {Pal}, \citenamefont {Wetli}, \citenamefont {Zamani}, \citenamefont {Stockert}, \citenamefont {L\"ohneysen}, \citenamefont {Fiebig},\ and\ \citenamefont {Kroha}}]{PhysRevLett.122.096401}%
  \BibitemOpen
  \bibfield  {author} {\bibinfo {author} {\bibfnamefont {S.}~\bibnamefont {Pal}}, \bibinfo {author} {\bibfnamefont {C.}~\bibnamefont {Wetli}}, \bibinfo {author} {\bibfnamefont {F.}~\bibnamefont {Zamani}}, \bibinfo {author} {\bibfnamefont {O.}~\bibnamefont {Stockert}}, \bibinfo {author} {\bibfnamefont {H.~v.}\ \bibnamefont {L\"ohneysen}}, \bibinfo {author} {\bibfnamefont {M.}~\bibnamefont {Fiebig}},\ and\ \bibinfo {author} {\bibfnamefont {J.}~\bibnamefont {Kroha}},\ }\bibfield  {title} {\bibinfo {title} {\textit{Fermi Volume Evolution and Crystal-Field Excitations in Heavy-Fermion Compounds Probed by Time-Domain Terahertz Spectroscopy}},\ }\href {https://doi.org/10.1103/PhysRevLett.122.096401} {\bibfield  {journal} {\bibinfo  {journal} {Phys. Rev. Lett.}\ }\textbf {\bibinfo {volume} {122}},\ \bibinfo {pages} {096401} (\bibinfo {year} {2019})}\BibitemShut {NoStop}%
\bibitem [{\citenamefont {Lawrence}\ \emph {et~al.}(1993)\citenamefont {Lawrence}, \citenamefont {Arko}, \citenamefont {Joyce}, \citenamefont {Blyth}, \citenamefont {Bartlett}, \citenamefont {Canfield}, \citenamefont {Fisk},\ and\ \citenamefont {Riseborough}}]{PhysRevB.47.15460}%
  \BibitemOpen
  \bibfield  {author} {\bibinfo {author} {\bibfnamefont {J.~M.}\ \bibnamefont {Lawrence}}, \bibinfo {author} {\bibfnamefont {A.~J.}\ \bibnamefont {Arko}}, \bibinfo {author} {\bibfnamefont {J.~J.}\ \bibnamefont {Joyce}}, \bibinfo {author} {\bibfnamefont {R.~I.~R.}\ \bibnamefont {Blyth}}, \bibinfo {author} {\bibfnamefont {R.~J.}\ \bibnamefont {Bartlett}}, \bibinfo {author} {\bibfnamefont {P.~C.}\ \bibnamefont {Canfield}}, \bibinfo {author} {\bibfnamefont {Z.}~\bibnamefont {Fisk}},\ and\ \bibinfo {author} {\bibfnamefont {P.~S.}\ \bibnamefont {Riseborough}},\ }\bibfield  {title} {\bibinfo {title} {\textit{Photoemission spectra of ${\mathrm{CeAl}}_{3}$, ${\mathrm{CeBe}}_{13}$, ${\mathrm{CeSi}}_{2}$, and ${\mathrm{CeCu}}_{2}$${\mathrm{Si}}_{2}$: Weights and widths of the 4f emission features}},\ }\href {https://doi.org/10.1103/PhysRevB.47.15460} {\bibfield  {journal} {\bibinfo  {journal} {Phys. Rev. B}\ }\textbf {\bibinfo {volume} {47}},\ \bibinfo {pages} {15460} (\bibinfo {year} {1993})}\BibitemShut {NoStop}%
\bibitem [{\citenamefont {Travlos}\ and\ \citenamefont {Salamouras}(1997)}]{TRAVLOS199713}%
  \BibitemOpen
  \bibfield  {author} {\bibinfo {author} {\bibfnamefont {A.}~\bibnamefont {Travlos}}\ and\ \bibinfo {author} {\bibfnamefont {N.}~\bibnamefont {Salamouras}},\ }\bibfield  {title} {\bibinfo {title} {\textit{Superconductivity of lanthanum silicide thin films}},\ }\href {https://doi.org/https://doi.org/10.1016/S0042-207X(96)00236-9} {\bibfield  {journal} {\bibinfo  {journal} {Vacuum}\ }\textbf {\bibinfo {volume} {48}},\ \bibinfo {pages} {13} (\bibinfo {year} {1997})}\BibitemShut {NoStop}%
\bibitem [{\citenamefont {Liu}\ \emph {et~al.}(2019)\citenamefont {Liu}, \citenamefont {Miao}, \citenamefont {Liu}, \citenamefont {Nan}, \citenamefont {Wang}, \citenamefont {Ge}, \citenamefont {Zhu}, \citenamefont {Yang}, \citenamefont {Wang},\ and\ \citenamefont {Guo}}]{PhysRevB.100.165308}%
  \BibitemOpen
  \bibfield  {author} {\bibinfo {author} {\bibfnamefont {L.}~\bibnamefont {Liu}}, \bibinfo {author} {\bibfnamefont {G.}~\bibnamefont {Miao}}, \bibinfo {author} {\bibfnamefont {B.}~\bibnamefont {Liu}}, \bibinfo {author} {\bibfnamefont {P.}~\bibnamefont {Nan}}, \bibinfo {author} {\bibfnamefont {Y.}~\bibnamefont {Wang}}, \bibinfo {author} {\bibfnamefont {B.}~\bibnamefont {Ge}}, \bibinfo {author} {\bibfnamefont {X.}~\bibnamefont {Zhu}}, \bibinfo {author} {\bibfnamefont {F.}~\bibnamefont {Yang}}, \bibinfo {author} {\bibfnamefont {W.}~\bibnamefont {Wang}},\ and\ \bibinfo {author} {\bibfnamefont {J.}~\bibnamefont {Guo}},\ }\bibfield  {title} {\bibinfo {title} {\textit{Interfacial effects on the superconducting properties of $\mathrm{LaS}{\mathrm{i}}_{2}(112)$ films on $\mathrm{Si(111)}$}},\ }\href {https://doi.org/10.1103/PhysRevB.100.165308} {\bibfield  {journal} {\bibinfo  {journal} {Phys. Rev. B}\ }\textbf {\bibinfo {volume} {100}},\ \bibinfo {pages} {165308} (\bibinfo {year} {2019})}\BibitemShut {NoStop}%
\bibitem [{\citenamefont {Cornut}\ and\ \citenamefont {Coqblin}(1972)}]{PhysRevB.5.4541}%
  \BibitemOpen
  \bibfield  {author} {\bibinfo {author} {\bibfnamefont {B.}~\bibnamefont {Cornut}}\ and\ \bibinfo {author} {\bibfnamefont {B.}~\bibnamefont {Coqblin}},\ }\bibfield  {title} {\bibinfo {title} {\textit{Influence of the Crystalline Field on the Kondo Effect of Alloys and Compounds with Cerium Impurities}},\ }\href {https://doi.org/10.1103/PhysRevB.5.4541} {\bibfield  {journal} {\bibinfo  {journal} {Phys. Rev. B}\ }\textbf {\bibinfo {volume} {5}},\ \bibinfo {pages} {4541} (\bibinfo {year} {1972})}\BibitemShut {NoStop}%
\bibitem [{\citenamefont {Stewart}(1984)}]{RevModPhys.56.755}%
  \BibitemOpen
  \bibfield  {author} {\bibinfo {author} {\bibfnamefont {G.~R.}\ \bibnamefont {Stewart}},\ }\bibfield  {title} {\bibinfo {title} {\textit{Heavy-fermion systems}},\ }\href {https://doi.org/10.1103/RevModPhys.56.755} {\bibfield  {journal} {\bibinfo  {journal} {Rev. Mod. Phys.}\ }\textbf {\bibinfo {volume} {56}},\ \bibinfo {pages} {755} (\bibinfo {year} {1984})}\BibitemShut {NoStop}%
\bibitem [{\citenamefont {K\"ohler}\ \emph {et~al.}(2008)\citenamefont {K\"ohler}, \citenamefont {Oeschler}, \citenamefont {Steglich}, \citenamefont {Maquilon},\ and\ \citenamefont {Fisk}}]{PhysRevB.77.104412}%
  \BibitemOpen
  \bibfield  {author} {\bibinfo {author} {\bibfnamefont {U.}~\bibnamefont {K\"ohler}}, \bibinfo {author} {\bibfnamefont {N.}~\bibnamefont {Oeschler}}, \bibinfo {author} {\bibfnamefont {F.}~\bibnamefont {Steglich}}, \bibinfo {author} {\bibfnamefont {S.}~\bibnamefont {Maquilon}},\ and\ \bibinfo {author} {\bibfnamefont {Z.}~\bibnamefont {Fisk}},\ }\bibfield  {title} {\bibinfo {title} {\textit{Energy scales of ${\mathrm{Lu}}_{1\ensuremath{-}x}{\mathrm{Yb}}_{x}{\mathrm{Rh}}_{2}{\mathrm{Si}}_{2}$ by means of thermopower investigations}},\ }\href {https://doi.org/10.1103/PhysRevB.77.104412} {\bibfield  {journal} {\bibinfo  {journal} {Phys. Rev. B}\ }\textbf {\bibinfo {volume} {77}},\ \bibinfo {pages} {104412} (\bibinfo {year} {2008})}\BibitemShut {NoStop}%
\bibitem [{\citenamefont {Ernst}\ \emph {et~al.}(2011)\citenamefont {Ernst}, \citenamefont {Kirchner}, \citenamefont {Krellner}, \citenamefont {Geibel}, \citenamefont {Zwicknagl}, \citenamefont {Steglich},\ and\ \citenamefont {Wirth}}]{Ernst2011}%
  \BibitemOpen
  \bibfield  {author} {\bibinfo {author} {\bibfnamefont {S.}~\bibnamefont {Ernst}}, \bibinfo {author} {\bibfnamefont {S.}~\bibnamefont {Kirchner}}, \bibinfo {author} {\bibfnamefont {C.}~\bibnamefont {Krellner}}, \bibinfo {author} {\bibfnamefont {C.}~\bibnamefont {Geibel}}, \bibinfo {author} {\bibfnamefont {G.}~\bibnamefont {Zwicknagl}}, \bibinfo {author} {\bibfnamefont {F.}~\bibnamefont {Steglich}},\ and\ \bibinfo {author} {\bibfnamefont {S.}~\bibnamefont {Wirth}},\ }\bibfield  {title} {\bibinfo {title} {\textit{Emerging local Kondo screening and spatial coherence in the heavy-fermion metal ${{\mathrm{YbRh}}_{2}{\mathrm{Si}}_{2}}$}},\ }\href {https://doi.org/10.1038/nature10148} {\bibfield  {journal} {\bibinfo  {journal} {Nature}\ }\textbf {\bibinfo {volume} {474}},\ \bibinfo {pages} {362} (\bibinfo {year} {2011})}\BibitemShut {NoStop}%
\bibitem [{\citenamefont {Pikul}\ \emph {et~al.}(2012)\citenamefont {Pikul}, \citenamefont {Stockert}, \citenamefont {Steppke}, \citenamefont {Cichorek}, \citenamefont {Hartmann}, \citenamefont {Caroca-Canales}, \citenamefont {Oeschler}, \citenamefont {Brando}, \citenamefont {Geibel},\ and\ \citenamefont {Steglich}}]{PhysRevLett.108.066405}%
  \BibitemOpen
  \bibfield  {author} {\bibinfo {author} {\bibfnamefont {A.~P.}\ \bibnamefont {Pikul}}, \bibinfo {author} {\bibfnamefont {U.}~\bibnamefont {Stockert}}, \bibinfo {author} {\bibfnamefont {A.}~\bibnamefont {Steppke}}, \bibinfo {author} {\bibfnamefont {T.}~\bibnamefont {Cichorek}}, \bibinfo {author} {\bibfnamefont {S.}~\bibnamefont {Hartmann}}, \bibinfo {author} {\bibfnamefont {N.}~\bibnamefont {Caroca-Canales}}, \bibinfo {author} {\bibfnamefont {N.}~\bibnamefont {Oeschler}}, \bibinfo {author} {\bibfnamefont {M.}~\bibnamefont {Brando}}, \bibinfo {author} {\bibfnamefont {C.}~\bibnamefont {Geibel}},\ and\ \bibinfo {author} {\bibfnamefont {F.}~\bibnamefont {Steglich}},\ }\bibfield  {title} {\bibinfo {title} {\textit{Single-Ion Kondo Scaling of the Coherent Fermi Liquid Regime in ${\mathrm{Ce}}_{1\ensuremath{-}x}{\mathrm{La}}_{x}{\mathrm{Ni}}_{2}{\mathrm{Ge}}_{2}$}},\ }\href {https://doi.org/10.1103/PhysRevLett.108.066405} {\bibfield  {journal} {\bibinfo  {journal} {Phys. Rev. Lett.}\ }\textbf {\bibinfo {volume} {108}},\ \bibinfo {pages} {066405} (\bibinfo {year} {2012})}\BibitemShut {NoStop}%
\bibitem [{\citenamefont {Brando}\ \emph {et~al.}(2016)\citenamefont {Brando}, \citenamefont {Belitz}, \citenamefont {Grosche},\ and\ \citenamefont {Kirkpatrick}}]{RevModPhys.88.025006}%
  \BibitemOpen
  \bibfield  {author} {\bibinfo {author} {\bibfnamefont {M.}~\bibnamefont {Brando}}, \bibinfo {author} {\bibfnamefont {D.}~\bibnamefont {Belitz}}, \bibinfo {author} {\bibfnamefont {F.~M.}\ \bibnamefont {Grosche}},\ and\ \bibinfo {author} {\bibfnamefont {T.~R.}\ \bibnamefont {Kirkpatrick}},\ }\bibfield  {title} {\bibinfo {title} {\textit{Metallic quantum ferromagnets}},\ }\href {https://doi.org/10.1103/RevModPhys.88.025006} {\bibfield  {journal} {\bibinfo  {journal} {Rev. Mod. Phys.}\ }\textbf {\bibinfo {volume} {88}},\ \bibinfo {pages} {025006} (\bibinfo {year} {2016})}\BibitemShut {NoStop}%
\bibitem [{\citenamefont {Steppke}\ \emph {et~al.}(2013)\citenamefont {Steppke}, \citenamefont {Küchler}, \citenamefont {Lausberg}, \citenamefont {Lengyel}, \citenamefont {Steinke}, \citenamefont {Borth}, \citenamefont {Lühmann}, \citenamefont {Krellner}, \citenamefont {Nicklas}, \citenamefont {Geibel}, \citenamefont {Steglich},\ and\ \citenamefont {Brando}}]{steppke2013science}%
  \BibitemOpen
  \bibfield  {author} {\bibinfo {author} {\bibfnamefont {A.}~\bibnamefont {Steppke}}, \bibinfo {author} {\bibfnamefont {R.}~\bibnamefont {Küchler}}, \bibinfo {author} {\bibfnamefont {S.}~\bibnamefont {Lausberg}}, \bibinfo {author} {\bibfnamefont {E.}~\bibnamefont {Lengyel}}, \bibinfo {author} {\bibfnamefont {L.}~\bibnamefont {Steinke}}, \bibinfo {author} {\bibfnamefont {R.}~\bibnamefont {Borth}}, \bibinfo {author} {\bibfnamefont {T.}~\bibnamefont {Lühmann}}, \bibinfo {author} {\bibfnamefont {C.}~\bibnamefont {Krellner}}, \bibinfo {author} {\bibfnamefont {M.}~\bibnamefont {Nicklas}}, \bibinfo {author} {\bibfnamefont {C.}~\bibnamefont {Geibel}}, \bibinfo {author} {\bibfnamefont {F.}~\bibnamefont {Steglich}},\ and\ \bibinfo {author} {\bibfnamefont {M.}~\bibnamefont {Brando}},\ }\bibfield  {title} {\bibinfo {title} {\textit{Ferromagnetic Quantum Critical Point in the Heavy-Fermion Metal ${\mathrm{Yb}}{\mathrm{Ni}}_{4}({\mathrm{P}}_{1-x}{\mathrm{As}}_{x})_{2}$ }},\ }\href {https://doi.org/10.1126/science.1230583} {\bibfield  {journal} {\bibinfo  {journal} {Science}\ }\textbf {\bibinfo {volume} {339}},\ \bibinfo {pages} {933} (\bibinfo {year} {2013})}\BibitemShut {NoStop}%
\bibitem [{\citenamefont {Shen}\ \emph {et~al.}(2020)\citenamefont {Shen}, \citenamefont {Zhang}, \citenamefont {Komijani}, \citenamefont {Nicklas}, \citenamefont {Borth}, \citenamefont {Wang}, \citenamefont {Chen}, \citenamefont {Nie}, \citenamefont {Li}, \citenamefont {Lu}, \citenamefont {Lee}, \citenamefont {Smidman}, \citenamefont {Steglich}, \citenamefont {Coleman},\ and\ \citenamefont {Yuan}}]{Shen2020}%
  \BibitemOpen
  \bibfield  {author} {\bibinfo {author} {\bibfnamefont {B.}~\bibnamefont {Shen}}, \bibinfo {author} {\bibfnamefont {Y.}~\bibnamefont {Zhang}}, \bibinfo {author} {\bibfnamefont {Y.}~\bibnamefont {Komijani}}, \bibinfo {author} {\bibfnamefont {M.}~\bibnamefont {Nicklas}}, \bibinfo {author} {\bibfnamefont {R.}~\bibnamefont {Borth}}, \bibinfo {author} {\bibfnamefont {A.}~\bibnamefont {Wang}}, \bibinfo {author} {\bibfnamefont {Y.}~\bibnamefont {Chen}}, \bibinfo {author} {\bibfnamefont {Z.}~\bibnamefont {Nie}}, \bibinfo {author} {\bibfnamefont {R.}~\bibnamefont {Li}}, \bibinfo {author} {\bibfnamefont {X.}~\bibnamefont {Lu}}, \bibinfo {author} {\bibfnamefont {H.}~\bibnamefont {Lee}}, \bibinfo {author} {\bibfnamefont {M.}~\bibnamefont {Smidman}}, \bibinfo {author} {\bibfnamefont {F.}~\bibnamefont {Steglich}}, \bibinfo {author} {\bibfnamefont {P.}~\bibnamefont {Coleman}},\ and\ \bibinfo {author} {\bibfnamefont {H.}~\bibnamefont {Yuan}},\ }\bibfield  {title} {\bibinfo {title} {\textit{Strange-metal behaviour in a pure ferromagnetic Kondo lattice}},\ }\href {https://doi.org/10.1038/s41586-020-2052-z} {\bibfield  {journal} {\bibinfo  {journal} {Nature}\ }\textbf {\bibinfo {volume} {579}},\ \bibinfo {pages} {51} (\bibinfo {year} {2020})}\BibitemShut {NoStop}%
\bibitem [{\citenamefont {Dai}\ \emph {et~al.}(2025)\citenamefont {Dai}, \citenamefont {Antezak}, \citenamefont {Broad}, \citenamefont {Thees}, \citenamefont {Zatko}, \citenamefont {Bouwmeester}, \citenamefont {Fortuna}, \citenamefont {Le~F\`evre}, \citenamefont {Rault}, \citenamefont {Horiba}, \citenamefont {Vyalikh}, \citenamefont {Kumigashira}, \citenamefont {Kliemt}, \citenamefont {Friedemann}, \citenamefont {Krellner}, \citenamefont {Frantzeskakis},\ and\ \citenamefont {Santander-Syro}}]{PhysRevLett.134.126401}%
  \BibitemOpen
  \bibfield  {author} {\bibinfo {author} {\bibfnamefont {J.}~\bibnamefont {Dai}}, \bibinfo {author} {\bibfnamefont {A.}~\bibnamefont {Antezak}}, \bibinfo {author} {\bibfnamefont {W.}~\bibnamefont {Broad}}, \bibinfo {author} {\bibfnamefont {M.}~\bibnamefont {Thees}}, \bibinfo {author} {\bibfnamefont {V.}~\bibnamefont {Zatko}}, \bibinfo {author} {\bibfnamefont {R.~L.}\ \bibnamefont {Bouwmeester}}, \bibinfo {author} {\bibfnamefont {F.}~\bibnamefont {Fortuna}}, \bibinfo {author} {\bibfnamefont {P.}~\bibnamefont {Le~F\`evre}}, \bibinfo {author} {\bibfnamefont {J.~E.}\ \bibnamefont {Rault}}, \bibinfo {author} {\bibfnamefont {K.}~\bibnamefont {Horiba}}, \bibinfo {author} {\bibfnamefont {D.~V.}\ \bibnamefont {Vyalikh}}, \bibinfo {author} {\bibfnamefont {H.}~\bibnamefont {Kumigashira}}, \bibinfo {author} {\bibfnamefont {K.}~\bibnamefont {Kliemt}}, \bibinfo {author} {\bibfnamefont {S.}~\bibnamefont {Friedemann}}, \bibinfo {author} {\bibfnamefont {C.}~\bibnamefont {Krellner}}, \bibinfo {author} {\bibfnamefont {E.}~\bibnamefont {Frantzeskakis}},\ and\ \bibinfo {author} {\bibfnamefont {A.~F.}\ \bibnamefont {Santander-Syro}},\ }\bibfield  {title} {\bibinfo {title} {\textit{Electronic Structure Dimensionality of the Quantum-Critical Ferromagnet ${\mathrm{YbNi}}_{4}{\mathrm{P}}_{2}$}},\ }\href {https://doi.org/10.1103/PhysRevLett.134.126401} {\bibfield  {journal} {\bibinfo  {journal} {Phys. Rev. Lett.}\ }\textbf {\bibinfo {volume} {134}},\ \bibinfo {pages} {126401} (\bibinfo {year} {2025})}\BibitemShut {NoStop}%
\bibitem [{\citenamefont {Aoki}\ \emph {et~al.}(1992)\citenamefont {Aoki}, \citenamefont {Yata}, \citenamefont {Isoda},\ and\ \citenamefont {Uji}}]{AOKI19921905}%
  \BibitemOpen
  \bibfield  {author} {\bibinfo {author} {\bibfnamefont {H.}~\bibnamefont {Aoki}}, \bibinfo {author} {\bibfnamefont {M.}~\bibnamefont {Yata}}, \bibinfo {author} {\bibfnamefont {Y.}~\bibnamefont {Isoda}},\ and\ \bibinfo {author} {\bibfnamefont {S.}~\bibnamefont {Uji}},\ }\bibfield  {title} {\bibinfo {title} {\textit{MBE growth of ${\mathrm{CeSi}_{2}}$ thin films and their electrical transport properties}},\ }\href {https://doi.org/https://doi.org/10.1016/0304-8853(92)91598-N} {\bibfield  {journal} {\bibinfo  {journal} {Journal of Magnetism and Magnetic Materials}\ }\textbf {\bibinfo {volume} {104-107}},\ \bibinfo {pages} {1905} (\bibinfo {year} {1992})}\BibitemShut {NoStop}%
\bibitem [{\citenamefont {Zheng}\ \emph {et~al.}(2025)\citenamefont {Zheng}, \citenamefont {Xiao}, \citenamefont {Pan}, \citenamefont {Yang}, \citenamefont {Liu}, \citenamefont {Bian}, \citenamefont {Wu}, \citenamefont {Hua}, \citenamefont {Zhang}, \citenamefont {Lu}, \citenamefont {Li}, \citenamefont {Sun}, \citenamefont {Song}, \citenamefont {He}, \citenamefont {Jim{\'e}nez}, \citenamefont {Cao}, \citenamefont {Yuan}, \citenamefont {Xu}, \citenamefont {Yin}, \citenamefont {Shi}, \citenamefont {Cao},\ and\ \citenamefont {Liu}}]{Zheng2025SCPMA}%
  \BibitemOpen
  \bibfield  {author} {\bibinfo {author} {\bibfnamefont {H.}~\bibnamefont {Zheng}}, \bibinfo {author} {\bibfnamefont {Z.}~\bibnamefont {Xiao}}, \bibinfo {author} {\bibfnamefont {Z.}~\bibnamefont {Pan}}, \bibinfo {author} {\bibfnamefont {G.}~\bibnamefont {Yang}}, \bibinfo {author} {\bibfnamefont {Y.}~\bibnamefont {Liu}}, \bibinfo {author} {\bibfnamefont {J.}~\bibnamefont {Bian}}, \bibinfo {author} {\bibfnamefont {Y.}~\bibnamefont {Wu}}, \bibinfo {author} {\bibfnamefont {T.}~\bibnamefont {Hua}}, \bibinfo {author} {\bibfnamefont {J.}~\bibnamefont {Zhang}}, \bibinfo {author} {\bibfnamefont {J.}~\bibnamefont {Lu}}, \bibinfo {author} {\bibfnamefont {J.}~\bibnamefont {Li}}, \bibinfo {author} {\bibfnamefont {T.}~\bibnamefont {Sun}}, \bibinfo {author} {\bibfnamefont {Y.}~\bibnamefont {Song}}, \bibinfo {author} {\bibfnamefont {R.}~\bibnamefont {He}}, \bibinfo {author} {\bibfnamefont {J.~L.}\ \bibnamefont {Jim{\'e}nez}}, \bibinfo {author} {\bibfnamefont {G.}~\bibnamefont {Cao}}, \bibinfo {author} {\bibfnamefont {H.}~\bibnamefont {Yuan}}, \bibinfo {author} {\bibfnamefont {Y.}~\bibnamefont {Xu}}, \bibinfo {author} {\bibfnamefont {Y.}~\bibnamefont {Yin}}, \bibinfo {author} {\bibfnamefont {M.}~\bibnamefont {Shi}}, \bibinfo {author} {\bibfnamefont {C.}~\bibnamefont {Cao}},\ and\ \bibinfo {author} {\bibfnamefont {Y.}~\bibnamefont {Liu}},\ }\bibfield  {title} {\bibinfo {title} {\textit{3d flat bands and coupled 4f moments in the kagome-honeycomb permanent magnet ${\mathrm{Sm}}_{2}{\mathrm{Co}}_{17}$}},\ }\href {https://doi.org/10.1007/s11433-025-2677-x} {\bibfield  {journal} {\bibinfo  {journal} {Science China Physics, Mechanics {\&} Astronomy}\ }\textbf {\bibinfo {volume} {68}},\ \bibinfo {pages} {287511} (\bibinfo {year} {2025})}\BibitemShut {NoStop}%
\bibitem [{\citenamefont {Kresse}\ and\ \citenamefont {Hafner}(1993)}]{PhysRevB.47.558}%
  \BibitemOpen
  \bibfield  {author} {\bibinfo {author} {\bibfnamefont {G.}~\bibnamefont {Kresse}}\ and\ \bibinfo {author} {\bibfnamefont {J.}~\bibnamefont {Hafner}},\ }\bibfield  {title} {\bibinfo {title} {\textit{Ab initio molecular dynamics for liquid metals}},\ }\href {https://doi.org/10.1103/PhysRevB.47.558} {\bibfield  {journal} {\bibinfo  {journal} {Phys. Rev. B}\ }\textbf {\bibinfo {volume} {47}},\ \bibinfo {pages} {558} (\bibinfo {year} {1993})}\BibitemShut {NoStop}%
\bibitem [{\citenamefont {Kresse}\ and\ \citenamefont {Joubert}(1999)}]{PhysRevB.59.1758}%
  \BibitemOpen
  \bibfield  {author} {\bibinfo {author} {\bibfnamefont {G.}~\bibnamefont {Kresse}}\ and\ \bibinfo {author} {\bibfnamefont {D.}~\bibnamefont {Joubert}},\ }\bibfield  {title} {\bibinfo {title} {\textit{From ultrasoft pseudopotentials to the projector augmented-wave method}},\ }\href {https://doi.org/10.1103/PhysRevB.59.1758} {\bibfield  {journal} {\bibinfo  {journal} {Phys. Rev. B}\ }\textbf {\bibinfo {volume} {59}},\ \bibinfo {pages} {1758} (\bibinfo {year} {1999})}\BibitemShut {NoStop}%
\bibitem [{\citenamefont {Perdew}\ \emph {et~al.}(1996)\citenamefont {Perdew}, \citenamefont {Burke},\ and\ \citenamefont {Ernzerhof}}]{PhysRevLett.77.3865}%
  \BibitemOpen
  \bibfield  {author} {\bibinfo {author} {\bibfnamefont {J.~P.}\ \bibnamefont {Perdew}}, \bibinfo {author} {\bibfnamefont {K.}~\bibnamefont {Burke}},\ and\ \bibinfo {author} {\bibfnamefont {M.}~\bibnamefont {Ernzerhof}},\ }\bibfield  {title} {\bibinfo {title} {\textit{Generalized Gradient Approximation Made Simple}},\ }\href {https://doi.org/10.1103/PhysRevLett.77.3865} {\bibfield  {journal} {\bibinfo  {journal} {Phys. Rev. Lett.}\ }\textbf {\bibinfo {volume} {77}},\ \bibinfo {pages} {3865} (\bibinfo {year} {1996})}\BibitemShut {NoStop}%
\bibitem [{\citenamefont {Wu}\ \emph {et~al.}(2021)\citenamefont {Wu}, \citenamefont {Zhang}, \citenamefont {Du}, \citenamefont {Shen}, \citenamefont {Zheng}, \citenamefont {Fang}, \citenamefont {Smidman}, \citenamefont {Cao}, \citenamefont {Steglich}, \citenamefont {Yuan}, \citenamefont {Denlinger},\ and\ \citenamefont {Liu}}]{PhysRevLett.126.216406}%
  \BibitemOpen
  \bibfield  {author} {\bibinfo {author} {\bibfnamefont {Y.}~\bibnamefont {Wu}}, \bibinfo {author} {\bibfnamefont {Y.}~\bibnamefont {Zhang}}, \bibinfo {author} {\bibfnamefont {F.}~\bibnamefont {Du}}, \bibinfo {author} {\bibfnamefont {B.}~\bibnamefont {Shen}}, \bibinfo {author} {\bibfnamefont {H.}~\bibnamefont {Zheng}}, \bibinfo {author} {\bibfnamefont {Y.}~\bibnamefont {Fang}}, \bibinfo {author} {\bibfnamefont {M.}~\bibnamefont {Smidman}}, \bibinfo {author} {\bibfnamefont {C.}~\bibnamefont {Cao}}, \bibinfo {author} {\bibfnamefont {F.}~\bibnamefont {Steglich}}, \bibinfo {author} {\bibfnamefont {H.}~\bibnamefont {Yuan}}, \bibinfo {author} {\bibfnamefont {J.~D.}\ \bibnamefont {Denlinger}},\ and\ \bibinfo {author} {\bibfnamefont {Y.}~\bibnamefont {Liu}},\ }\bibfield  {title} {\bibinfo {title} {\textit{Anisotropic $c\ensuremath{-}f$ Hybridization in the Ferromagnetic Quantum Critical Metal ${\mathrm{CeRh}}_{6}{\mathrm{Ge}}_{4}$}},\ }\href {https://doi.org/10.1103/PhysRevLett.126.216406} {\bibfield  {journal} {\bibinfo  {journal} {Phys. Rev. Lett.}\ }\textbf {\bibinfo {volume} {126}},\ \bibinfo {pages} {216406} (\bibinfo {year} {2021})}\BibitemShut {NoStop}%
\bibitem [{\citenamefont {Wu}\ \emph {et~al.}(2024)\citenamefont {Wu}, \citenamefont {Zhang}, \citenamefont {Ju}, \citenamefont {Hu}, \citenamefont {Huang}, \citenamefont {Zhang}, \citenamefont {Zhang}, \citenamefont {Zheng}, \citenamefont {Yang}, \citenamefont {Eljaouhari}, \citenamefont {Song}, \citenamefont {C.~Plumb}, \citenamefont {Steglich}, \citenamefont {Shi}, \citenamefont {Zwicknagl}, \citenamefont {Cao}, \citenamefont {Yuan},\ and\ \citenamefont {Liu}}]{Wu_2024}%
  \BibitemOpen
  \bibfield  {author} {\bibinfo {author} {\bibfnamefont {Y.}~\bibnamefont {Wu}}, \bibinfo {author} {\bibfnamefont {Y.}~\bibnamefont {Zhang}}, \bibinfo {author} {\bibfnamefont {S.}~\bibnamefont {Ju}}, \bibinfo {author} {\bibfnamefont {Y.}~\bibnamefont {Hu}}, \bibinfo {author} {\bibfnamefont {Y.}~\bibnamefont {Huang}}, \bibinfo {author} {\bibfnamefont {Y.}~\bibnamefont {Zhang}}, \bibinfo {author} {\bibfnamefont {H.}~\bibnamefont {Zhang}}, \bibinfo {author} {\bibfnamefont {H.}~\bibnamefont {Zheng}}, \bibinfo {author} {\bibfnamefont {G.}~\bibnamefont {Yang}}, \bibinfo {author} {\bibfnamefont {E.-O.}\ \bibnamefont {Eljaouhari}}, \bibinfo {author} {\bibfnamefont {B.}~\bibnamefont {Song}}, \bibinfo {author} {\bibfnamefont {N.}~\bibnamefont {C.~Plumb}}, \bibinfo {author} {\bibfnamefont {F.}~\bibnamefont {Steglich}}, \bibinfo {author} {\bibfnamefont {M.}~\bibnamefont {Shi}}, \bibinfo {author} {\bibfnamefont {G.}~\bibnamefont {Zwicknagl}}, \bibinfo {author} {\bibfnamefont {C.}~\bibnamefont {Cao}}, \bibinfo {author} {\bibfnamefont {H.}~\bibnamefont {Yuan}},\ and\ \bibinfo {author} {\bibfnamefont {Y.}~\bibnamefont {Liu}},\ }\bibfield  {title} {\bibinfo {title} {\textit{Fermi Surface Nesting with Heavy Quasiparticles in the Locally Noncentrosymmetric Superconductor ${\mathrm{CeRh}}_{2}{\mathrm{As}}_{2}$}},\ }\href {https://doi.org/10.1088/0256-307X/41/9/097403} {\bibfield  {journal} {\bibinfo  {journal} {Chinese Physics Letters}\ }\textbf {\bibinfo {volume} {41}},\ \bibinfo {pages} {097403} (\bibinfo {year} {2024})}\BibitemShut {NoStop}%
\bibitem [{\citenamefont {Mostofi}\ \emph {et~al.}(2008)\citenamefont {Mostofi}, \citenamefont {Yates}, \citenamefont {Lee}, \citenamefont {Souza}, \citenamefont {Vanderbilt},\ and\ \citenamefont {Marzari}}]{MOSTOFI2008685}%
  \BibitemOpen
  \bibfield  {author} {\bibinfo {author} {\bibfnamefont {A.~A.}\ \bibnamefont {Mostofi}}, \bibinfo {author} {\bibfnamefont {J.~R.}\ \bibnamefont {Yates}}, \bibinfo {author} {\bibfnamefont {Y.-S.}\ \bibnamefont {Lee}}, \bibinfo {author} {\bibfnamefont {I.}~\bibnamefont {Souza}}, \bibinfo {author} {\bibfnamefont {D.}~\bibnamefont {Vanderbilt}},\ and\ \bibinfo {author} {\bibfnamefont {N.}~\bibnamefont {Marzari}},\ }\bibfield  {title} {\bibinfo {title} {\textit{wannier90: A tool for obtaining maximally-localised Wannier functions}},\ }\href {https://doi.org/https://doi.org/10.1016/j.cpc.2007.11.016} {\bibfield  {journal} {\bibinfo  {journal} {Computer Physics Communications}\ }\textbf {\bibinfo {volume} {178}},\ \bibinfo {pages} {685 } (\bibinfo {year} {2008})}\BibitemShut {NoStop}%
\bibitem [{\citenamefont {Miccoli}\ \emph {et~al.}(2015)\citenamefont {Miccoli}, \citenamefont {Edler}, \citenamefont {Pfnür},\ and\ \citenamefont {Tegenkamp}}]{Miccoli_2015}%
  \BibitemOpen
  \bibfield  {author} {\bibinfo {author} {\bibfnamefont {I.}~\bibnamefont {Miccoli}}, \bibinfo {author} {\bibfnamefont {F.}~\bibnamefont {Edler}}, \bibinfo {author} {\bibfnamefont {H.}~\bibnamefont {Pfnür}},\ and\ \bibinfo {author} {\bibfnamefont {C.}~\bibnamefont {Tegenkamp}},\ }\bibfield  {title} {\bibinfo {title} {The 100th anniversary of the four-point probe technique: the role of probe geometries in isotropic and anisotropic systems},\ }\href {https://doi.org/10.1088/0953-8984/27/22/223201} {\bibfield  {journal} {\bibinfo  {journal} {Journal of Physics: Condensed Matter}\ }\textbf {\bibinfo {volume} {27}},\ \bibinfo {pages} {223201} (\bibinfo {year} {2015})}\BibitemShut {NoStop}%
\bibitem [{\citenamefont {Wu}\ \emph {et~al.}()\citenamefont {Wu}, \citenamefont {Zhu}, \citenamefont {Hua}, \citenamefont {Fang}, \citenamefont {Zhang}, \citenamefont {Zhang}, \citenamefont {Huang}, \citenamefont {Zheng}, \citenamefont {Fu}, \citenamefont {Zheng}, \citenamefont {Liu}, \citenamefont {Ye}, \citenamefont {Chen}, \citenamefont {Sun}, \citenamefont {Smidman}, \citenamefont {Kroha}, \citenamefont {Cao}, \citenamefont {Yuan}, \citenamefont {Steglich}, \citenamefont {Lin},\ and\ \citenamefont {Liu}}]{CeSi2_figshare}%
  \BibitemOpen
  \bibfield  {author} {\bibinfo {author} {\bibfnamefont {Y.}~\bibnamefont {Wu}}, \bibinfo {author} {\bibfnamefont {W.}~\bibnamefont {Zhu}}, \bibinfo {author} {\bibfnamefont {T.}~\bibnamefont {Hua}}, \bibinfo {author} {\bibfnamefont {Y.}~\bibnamefont {Fang}}, \bibinfo {author} {\bibfnamefont {Y.}~\bibnamefont {Zhang}}, \bibinfo {author} {\bibfnamefont {J.}~\bibnamefont {Zhang}}, \bibinfo {author} {\bibfnamefont {Y.}~\bibnamefont {Huang}}, \bibinfo {author} {\bibfnamefont {H.}~\bibnamefont {Zheng}}, \bibinfo {author} {\bibfnamefont {S.}~\bibnamefont {Fu}}, \bibinfo {author} {\bibfnamefont {X.}~\bibnamefont {Zheng}}, \bibinfo {author} {\bibfnamefont {Z.}~\bibnamefont {Liu}}, \bibinfo {author} {\bibfnamefont {M.}~\bibnamefont {Ye}}, \bibinfo {author} {\bibfnamefont {Y.}~\bibnamefont {Chen}}, \bibinfo {author} {\bibfnamefont {T.}~\bibnamefont {Sun}}, \bibinfo {author} {\bibfnamefont {M.}~\bibnamefont {Smidman}}, \bibinfo {author} {\bibfnamefont {J.}~\bibnamefont {Kroha}}, \bibinfo {author} {\bibfnamefont {C.}~\bibnamefont {Cao}}, \bibinfo {author} {\bibfnamefont {H.}~\bibnamefont {Yuan}}, \bibinfo {author} {\bibfnamefont {F.}~\bibnamefont {Steglich}}, \bibinfo {author} {\bibfnamefont {H.-Q.}\ \bibnamefont {Lin}},\ and\ \bibinfo {author} {\bibfnamefont {Y.}~\bibnamefont {Liu}},\ }\href@noop {} {\bibinfo {title} {\textit{Original data for {``Dimensionality tuning of heavy-fermion states in ultrathin ${\mathrm{CeSi}}_{2}$ films''}}}},\ \bibinfo {howpublished} {figshare repository, \url{https://doi.org/10.6084/m9.figshare.31286731}}\BibitemShut {NoStop}%
\end{thebibliography}
\end{document}